\documentclass[10pt, conference, letterpaper]{IEEEtran}
\usepackage[pass]{geometry}

\usepackage{algorithm}
\usepackage{amsfonts}
\usepackage{amsthm}
\usepackage{amsmath}
\usepackage{mathtools}
\usepackage{amssymb}
\usepackage{amsfonts,color,amsxtra} 
\usepackage{array}
\usepackage{bbm}
\usepackage{cite}
\usepackage{dsfont}
\usepackage{epsfig}
\usepackage{float}
\usepackage[T1]{fontenc}
\usepackage{graphicx}
\usepackage{indentfirst}
\usepackage{multicol}
\usepackage{subcaption}
\usepackage[normalem]{ulem} % to strike out the text
\usepackage{blindtext}
\usepackage[noend]{algpseudocode}
\usepackage[table]{xcolor}% http://ctan.org/pkg/xcolor
\usepackage{color, colortbl}
\usepackage{enumerate}
\usepackage{graphicx}

\usepackage{url}
\usepackage{multirow}
\usepackage{color}
\usepackage{blkarray} %added for matrix
\usepackage{tikz}%added to color  matrix element
\usetikzlibrary{calc}%added to color  matrix element
\usepackage[export]{adjustbox} % top alignment of subfigures
\usepackage{setspace}%for line spacing
\usepackage{comment}
\usepackage{algorithmicx}
\usepackage{algpseudocode}

\usepackage{etex}
\usepackage[para]{footmisc} % horizontal footnote

\usepackage{tikz}

\usepackage{xcolor}
\makeatletter
\def\@hex@@Hex#1%
 {\if a#1A\else \if b#1B\else \if c#1C\else \if d#1D\else
  \if e#1E\else \if f#1F\else #1\fi\fi\fi\fi\fi\fi \@hex@Hex}
\makeatother
\definecolor{qcolor0}{HTML}{8dd3c7}	% green
\definecolor{qcolor1}{HTML}{ffffb3}	% yellow
\definecolor{qcolor2}{HTML}{bebada}	% purple
\definecolor{qcolor3}{HTML}{fb8072}	% read
\definecolor{qcolor4}{HTML}{80b1d3}	% blue
\definecolor{qcolor5}{HTML}{fdb462}	% orange
\definecolor{qcolor6}{HTML}{b3de69}	% bright green
\definecolor{qcolor7}{HTML}{fccde5}	% pink
\definecolor{qcolor8}{HTML}{d9d9d9}	% gray
\definecolor{qcolor9}{HTML}{bc80bd}	% dark purple 
\definecolor{qcolor10}{HTML}{ccebc5} % light green
\definecolor{qcolor11}{HTML}{ffed6f} % dark yellow

\definecolor{gcolor0}{HTML}{ffffff}
\definecolor{gcolor1}{HTML}{deffff}
\definecolor{gcolor2}{HTML}{a4ccd3}
\definecolor{gcolor3}{HTML}{6d9ba9}
\definecolor{gcolor4}{HTML}{3a6b82}
\definecolor{gcolor5}{HTML}{003f5c}

\definecolor{myblue}{HTML}{396AB1} % blue
\definecolor{myorange}{HTML}{DA7C30} % orange
\definecolor{mygreen}{HTML}{3E9651}
\definecolor{myred}{HTML}{CC2529}
\definecolor{mygray}{HTML}{535154}
\definecolor{mypurple}{HTML}{6B4C9A}
\definecolor{mymaroon}{HTML}{922428}
\definecolor{mymudgreen}{HTML}{948B3D}

\definecolor{mylblue}{HTML}{7293CB} % blue
\definecolor{mylorange}{HTML}{E1974C} % orange
\definecolor{mylgreen}{HTML}{84BA5B}
\definecolor{mylred}{HTML}{D35E60}
\definecolor{mylgray}{HTML}{535154}
\definecolor{mylpurple}{HTML}{6B4C9A}
\definecolor{mylmaroon}{HTML}{922428}
\definecolor{mylmudgreen}{HTML}{948B3D}

\algrenewcommand\algorithmicforall{\textbf{foreach}}
\algrenewcommand\algorithmicindent{.8em}

\newcommand*\circled[1]{\tikz[baseline=(char.base)]{
            \node[shape=circle,draw,inner sep=1pt] (char) {#1};}}
            
\DeclareRobustCommand\circled[1]{\tikz[baseline=(char.base)]{
            \node[shape=circle,draw,inner sep=1pt] (char) {#1};}}

\usepackage{array}
\newcolumntype{L}[1]{>{\raggedright\let\newline\\\arraybackslash\hspace{0pt}}m{#1}}
\newcolumntype{C}[1]{>{\centering\let\newline\\\arraybackslash\hspace{0pt}}m{#1}}
\newcolumntype{R}[1]{>{\raggedleft\let\newline\\\arraybackslash\hspace{0pt}}m{#1}}

\usepackage{todonotes}%disable todo notes: [disable]

\DeclareMathSizes{10pt}{9pt}{6pt}{5pt} 	% Reduce math font size


\definecolor{Gray}{gray}{0.9}
\definecolor{LightCyan}{rgb}{0.9,0.9,0.9}

\newcommand{\fref}[1]{Fig.~\ref{#1}}

\newcommand{\aref}[1]{Algorithm~\ref{#1}}
\newcommand{\sref}[1]{Section~\ref{#1}}
\newcommand{\tref}[1]{Table~\ref{#1}}

\newenvironment{mathalign}{\par\nobreak\noindent\addtolength{\jot}{-0.2em}\align}{\endalign}
\newfont{\mycrnotice}{ptmr8t at 7pt}
\newfont{\myconfname}{ptmri8t at 7pt}

\IEEEoverridecommandlockouts

\begin{document}

\def\sharedaffiliation{%
\end{tabular}
\begin{tabular}{c}}

\title{Joint Relaying and Spatial Sharing Multicast Scheduling for mmWave Networks}

\author{
    \IEEEauthorblockN{Gek Hong (Allyson) Sim\IEEEauthorrefmark{1}, Mahdi Mousavi\IEEEauthorrefmark{2}, Lin Wang\IEEEauthorrefmark{3}, Anja Klein\IEEEauthorrefmark{2}, Matthias Hollick\IEEEauthorrefmark{1}}
    \IEEEauthorblockA{\IEEEauthorrefmark{1} Secure Mobile Networking Lab (SEEMOO), Technische Universit\"{a}t Darmstadt
    \\\{asim, mhollick\}@seemoo.tu-darmstadt.de}
    \IEEEauthorblockA{\IEEEauthorrefmark{2} Communication Engineering Lab, Technische Universit\"{a}t Darmstadt
    \\\{m.mousavi, a.klein\}@nt.tu-darmstadt.de}
    \IEEEauthorblockA{\IEEEauthorrefmark{3}VU Amsterdam, The Netherlands
    \\lin.wang@vu.nl}
}

\maketitle

\begin{abstract}
Millimeter-wave (mmWave) communication plays a vital role to efficiently disseminate large volumes of data in beyond-5G networks. 
Unfortunately, the directionality of mmWave communication significantly complicates efficient data dissemination, particularly in multicasting, which is gaining more and more importance in emerging applications (e.g., V2X, public safety). While multicasting for systems operating at lower frequencies (i.e., sub-6GHz) has been extensively studied, they are sub-optimal for mmWave systems as mmWave has significantly different propagation characteristics, i.e., using the directional transmission to compensate for the high path loss and thus promoting spectrum sharing.  
In this paper, we propose novel multicast scheduling algorithms by jointly exploiting relaying and spatial sharing gains while aiming to minimize the multicast completion time. We first characterize the min-time mmWave multicasting problem with a comprehensive model and formulate it with an integer linear program (ILP). We further design a practical and scalable distributed algorithm named \texttt{mmDiMu}, based on gradually maximizing the transmission throughput over time. Finally, we carry out validation through extensive simulations in different scales and the results show that \texttt{mmDiMu} significantly outperforms conventional algorithms with around $95\%$ reduction on multicast completion time.
\end{abstract}


\section{Introduction}
\label{sec:intro}

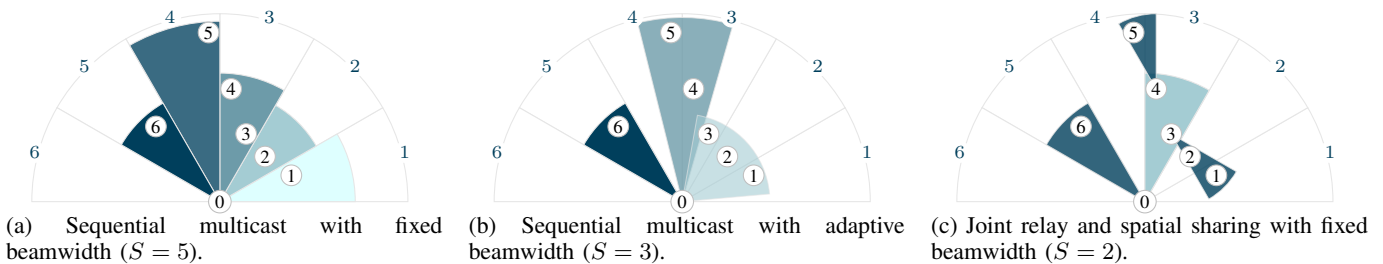
\begin{figure*}[t!]
\centering

	\begin{subfigure}{0.32\textwidth}
	\centering
	\begin{tikzpicture}
		\filldraw[fill=gcolor0, draw=gray!20] (0,0) -- (25mm,0mm) arc (0:30:25mm) -- cycle;
		\filldraw[fill=gcolor0, draw=gray!20] (0,0) -- (21.65mm,12.5mm) arc (30:60:25mm) -- cycle;
		\filldraw[fill=gcolor0, draw=gray!20] (0,0) -- (12.5mm,21.65mm) arc (60:90:25mm) -- cycle;
		\filldraw[fill=gcolor0, draw=gray!20] (0,0) -- (0mm,25mm) arc (90:120:25mm) -- cycle;
		\filldraw[fill=gcolor0, draw=gray!20] (0,0) -- (-12.5mm,21.65mm) arc (120:150:25mm) -- cycle;
		\filldraw[fill=gcolor0, draw=gray!20] (0,0) -- (-21.65mm,12.5mm) arc (150:180:25mm) -- cycle;
		\filldraw[fill=gcolor1, draw=gray!20] (0,0) -- (18mm,0mm) arc (0:30:18mm) -- cycle;
		\filldraw[fill=gcolor2, draw=gray!20] (0,0) -- (12.82mm,7.4mm) arc (30:60:14.81mm) -- cycle;
		\filldraw[fill=gcolor3, draw=gray!20] (0,0) -- (8.55mm,14.81mm) arc (60:90:17.1mm) -- cycle;
		\filldraw[fill=gcolor4, draw=gray!20] (0,0) -- (0mm,24mm) arc (90:120:24mm) -- cycle;
		\filldraw[fill=gcolor5, draw=gray!20] (0,0) -- (-7.56mm,13.1mm) arc (120:150:15.12mm) -- cycle;
%		\filldraw[fill=gcolor0, draw=gray!20] (0,0) -- (-21.65mm,12.5mm) arc (150:180:25mm) -- cycle;
		\node at (0, 0) [circle, inner sep=0pt, minimum size=3mm, draw=gray!50, fill=white, font=\scriptsize]{0}; 
		\node at (0.95, 0.35) [circle, inner sep=0pt, minimum size=2.75mm, draw=gray!50, fill=white, font=\scriptsize]{1};
		\node at (0.6, 0.6) [circle, inner sep=0pt, minimum size=2.75mm, draw=gray!50, fill=white, font=\scriptsize]{2}; 
		\node at (0.35, 0.9) [circle, inner sep=0pt, minimum size=2.75mm, draw=gray!50, fill=white, font=\scriptsize]{3};
		\node at (0.15, 1.5) [circle, inner sep=0pt, minimum size=2.75mm, draw=gray!50, fill=white, font=\scriptsize]{4};
		\node at (-0.15, 2.25) [circle, inner sep=0pt, minimum size=2.75mm, draw=gray!50, fill=white, font=\scriptsize]{5};
		\node at (-0.85, 1) [circle, inner sep=0pt, minimum size=2.75mm, draw=gray!50, fill=white, font=\scriptsize]{6};
		\node at (2.46, 0.66) [circle, inner sep=0pt, minimum size=2.75mm, draw=gray!0, fill=white, font=\scriptsize, text=gcolor5]{$1$};
		\node at (1.80, 1.80) [circle, inner sep=0pt, minimum size=2.75mm, draw=gray!0, fill=white, font=\scriptsize, text=gcolor5]{$2$};
		\node at (0.66, 2.46) [circle, inner sep=0pt, minimum size=2.75mm, draw=gray!0, fill=white, font=\scriptsize, text=gcolor5]{$3$};
		\node at (-0.66, 2.46) [circle, inner sep=0pt, minimum size=2.75mm, draw=gray!0, fill=white, font=\scriptsize, text=gcolor5]{$4$};
		\node at (-1.80, 1.80) [circle, inner sep=0pt, minimum size=2.75mm, draw=gray!0, fill=white, font=\scriptsize, text=gcolor5]{$5$};
		\node at (-2.46, 0.66) [circle, inner sep=0pt, minimum size=2.75mm, draw=gray!0, fill=white, font=\scriptsize, text=gcolor5]{$6$};
	\end{tikzpicture}
	\vspace{-2mm}
 	\caption{Sequential multicast with fixed beamwidth ($S = 5$).}
	\label{fig:sweep}
	\end{subfigure}
	\hspace{1mm}
	\begin{subfigure}{0.32\textwidth}
	\centering
	\begin{tikzpicture}
		\filldraw[fill=gcolor0, draw=gray!20] (0,0) -- (25mm,0mm) arc (0:30:25mm) -- cycle;
		\filldraw[fill=gcolor0, draw=gray!20] (0,0) -- (21.65mm,12.5mm) arc (30:60:25mm) -- cycle;
		\filldraw[fill=gcolor0, draw=gray!20] (0,0) -- (12.5mm,21.65mm) arc (60:90:25mm) -- cycle;
		\filldraw[fill=gcolor0, draw=gray!20] (0,0) -- (0mm,25mm) arc (90:120:25mm) -- cycle;
		\filldraw[fill=gcolor0, draw=gray!20] (0,0) -- (-12.5mm,21.65mm) arc (120:150:25mm) -- cycle;
		\filldraw[fill=gcolor0, draw=gray!20] (0,0) -- (-21.65mm,12.5mm) arc (150:180:25mm) -- cycle;
		\filldraw[fill=gcolor3, draw=gray!20, opacity=0.8] (0,0) -- (6.75mm,23.55mm) arc (74:104:24.5mm) -- cycle;
		\filldraw[fill=gcolor2, draw=gray!20, opacity=0.6] (0,0) -- (11.61mm, 1.02mm) arc (5:80:11.66mm) -- cycle;
		%\filldraw[fill=gcolor5, draw=gray!20] (0,0) -- (-8.01mm,12.82mm) arc (122:138:15.12mm) -- cycle;
		\filldraw[fill=gcolor5, draw=gray!20] (0,0) -- (-7.56mm,13.1mm) arc (120:150:15.12mm) -- cycle;
		\node at (0, 0) [circle, inner sep=0pt, minimum size=3mm, draw=gray!50, fill=white, font=\scriptsize]{0}; 
		\node at (0.95, 0.35) [circle, inner sep=0pt, minimum size=2.75mm, draw=gray!50, fill=white, font=\scriptsize]{1};
		\node at (0.6, 0.6) [circle, inner sep=0pt, minimum size=2.75mm, draw=gray!50, fill=white, font=\scriptsize]{2}; 
		\node at (0.35, 0.9) [circle, inner sep=0pt, minimum size=2.75mm, draw=gray!50, fill=white, font=\scriptsize]{3};
		\node at (0.15, 1.5) [circle, inner sep=0pt, minimum size=2.75mm, draw=gray!50, fill=white, font=\scriptsize]{4};
		\node at (-0.15, 2.25) [circle, inner sep=0pt, minimum size=2.75mm, draw=gray!50, fill=white, font=\scriptsize]{5};
		\node at (-0.85, 1) [circle, inner sep=0pt, minimum size=2.75mm, draw=gray!50, fill=white, font=\scriptsize]{6};
		\node at (2.46, 0.66) [circle, inner sep=0pt, minimum size=2.75mm, draw=gray!0, fill=white, font=\scriptsize, text=gcolor5]{$1$};
		\node at (1.80, 1.80) [circle, inner sep=0pt, minimum size=2.75mm, draw=gray!0, fill=white, font=\scriptsize, text=gcolor5]{$2$};
		\node at (0.66, 2.46) [circle, inner sep=0pt, minimum size=2.75mm, draw=gray!0, fill=white, font=\scriptsize, text=gcolor5]{$3$};
		\node at (-0.66, 2.46) [circle, inner sep=0pt, minimum size=2.75mm, draw=gray!0, fill=white, font=\scriptsize, text=gcolor5]{$4$};
		\node at (-1.80, 1.80) [circle, inner sep=0pt, minimum size=2.75mm, draw=gray!0, fill=white, font=\scriptsize, text=gcolor5]{$5$};
		\node at (-2.46, 0.66) [circle, inner sep=0pt, minimum size=2.75mm, draw=gray!0, fill=white, font=\scriptsize, text=gcolor5]{$6$};
	\end{tikzpicture}
 	\vspace{-2mm}
	\caption{Sequential multicast with adaptive beamwidth ($S = 3$).}
	\label{fig:hierarchical}
	\end{subfigure}
	\hspace{1mm}
	\begin{subfigure}{0.32\textwidth}
	\centering
	\begin{tikzpicture}
		\filldraw[fill=gcolor0, draw=gray!20] (0,0) -- (25mm,0mm) arc (0:30:25mm) -- cycle;
		\filldraw[fill=gcolor0, draw=gray!20] (0,0) -- (21.65mm,12.5mm) arc (30:60:25mm) -- cycle;
		\filldraw[fill=gcolor0, draw=gray!20] (0,0) -- (12.5mm,21.65mm) arc (60:90:25mm) -- cycle;
		\filldraw[fill=gcolor0, draw=gray!20] (0,0) -- (0mm,25mm) arc (90:120:25mm) -- cycle;
		\filldraw[fill=gcolor0, draw=gray!20] (0,0) -- (-12.5mm,21.65mm) arc (120:150:25mm) -- cycle;
		\filldraw[fill=gcolor0, draw=gray!20] (0,0) -- (-21.65mm,12.5mm) arc (150:180:25mm) -- cycle;
		\filldraw[fill=gcolor2, draw=gray!20] (0,0) -- (8.55mm,14.81mm) arc (60:90:17.1mm) -- cycle; 		% sector 3
		\filldraw[fill=gcolor5, draw=gray!20, opacity=0.8] (0.35,0.9) -- (8.5mm,0.34mm) arc (300:330:10mm) -- cycle; 	% sector 3
		\filldraw[fill=gcolor5, draw=gray!20, opacity=0.8] (0.15,1.5) -- (1.5mm,25mm) arc (90:120:10mm) -- cycle; 		% sector 3
		\filldraw[fill=gcolor5, draw=gray!20, opacity=0.8] (0,0) -- (-7.56mm,13.1mm) arc (120:150:15.12mm) -- cycle; 	% sector 5
		\node at (0, 0) [circle, inner sep=0pt, minimum size=3mm, draw=gray!50, fill=white, font=\scriptsize]{0};
		\node at (0.95, 0.35) [circle, inner sep=0pt, minimum size=2.75mm, draw=gray!50, fill=white, font=\scriptsize]{1};
		\node at (0.6, 0.6) [circle, inner sep=0pt, minimum size=2.75mm, draw=gray!50, fill=white, font=\scriptsize]{2}; 
		\node at (0.35, 0.9) [circle, inner sep=0pt, minimum size=2.75mm, draw=gray!50, fill=white, font=\scriptsize]{3};
		\node at (0.15, 1.5) [circle, inner sep=0pt, minimum size=2.75mm, draw=gray!50, fill=white, font=\scriptsize]{4};
		\node at (-0.15, 2.25) [circle, inner sep=0pt, minimum size=2.75mm, draw=gray!50, fill=white, font=\scriptsize]{5};
		\node at (-0.85, 1) [circle, inner sep=0pt, minimum size=2.75mm, draw=gray!50, fill=white, font=\scriptsize]{6};
		\node at (2.46, 0.66) [circle, inner sep=0pt, minimum size=2.75mm, draw=gray!0, fill=white, font=\scriptsize, text=gcolor5]{$1$};
		\node at (1.80, 1.80) [circle, inner sep=0pt, minimum size=2.75mm, draw=gray!0, fill=white, font=\scriptsize, text=gcolor5]{$2$};
		\node at (0.66, 2.46) [circle, inner sep=0pt, minimum size=2.75mm, draw=gray!0, fill=white, font=\scriptsize, text=gcolor5]{$3$};
		\node at (-0.66, 2.46) [circle, inner sep=0pt, minimum size=2.75mm, draw=gray!0, fill=white, font=\scriptsize, text=gcolor5]{$4$};
		\node at (-1.80, 1.80) [circle, inner sep=0pt, minimum size=2.75mm, draw=gray!0, fill=white, font=\scriptsize, text=gcolor5]{$5$};
		\node at (-2.46, 0.66) [circle, inner sep=0pt, minimum size=2.75mm, draw=gray!0, fill=white, font=\scriptsize, text=gcolor5]{$6$};
	\end{tikzpicture}
	\vspace{-2mm}
 	\caption{Joint relay and spatial sharing with fixed beamwidth ($S = 2$).}
	\label{fig:relay}
	\end{subfigure}
	\vspace{-2mm}
	\caption{Different multicast mechanisms with different gray shades represent the transmission/multicast sessions in different time slots, and $S$ indicates the total slots. The number at each sector's arc indicate the index of each sector. }
	\label{fig:multicast_scenarios}
	\vspace{-7mm}
\end{figure*}

Millimeter-wave (mmWave) communication fulfills the demand for multi-gigabit-per-second (Gbps) throughput and low-latency communication even for extremely dense networks~\cite{xiao:2017:jsac}, which are usually not easy to sustain with traditional communications operating at sub-6GHz frequencies. Despite its benefits, mmWave communication suffers from very high attenuation, resulting in dramatic penetration loss, due to its high frequency. To compensate for this loss, directional transmissions are typically employed, where the coverage of communication is constrained to a rather small area, e.g., to the line of sight in the extreme case. This limitation poses new challenges in particular to guarantee efficient content dissemination for various delay-sensitive multicast applications (e.g., raw sensory data broadcasting in vehicle-to-everything (V2X) communications to support autonomous driving, high-definition video broadcasting in a concert hall, and public-safety use cases). 
 
Although multicast scheduling has been widely explored for networks operating at sub-6GHz frequencies, the specific benefits and challenges of mmWave multicast scheduling remain understudied \cite{biason:2019:multicast}. In particular, multicast scheduler designs for sub-6GHz communications assume the availability of omnidirectional transmission, and thus a source node can schedule the transmission to any arbitrary subset of receiving nodes within a certain range simultaneously. However, the restricted coverage of mmWave communication undermines this assumption and renders these designs inapplicable, opening a new research question.

One trivial design for mmWave multicast scheduling can simply employ multiple directional unicast and/or multicast transmissions to sequentially serve all multicast nodes. The behavior of such a scheduler is illustrated in \fref{fig:sweep}, where the source node (labeled as {\footnotesize\circled{0}}) transmits sequentially in sectors $1$ to $5$ to serve multicast nodes {\footnotesize\circled{1}}, {\footnotesize\circled{2}}, {\footnotesize\circled{3} \circled{4}}, {\footnotesize\circled{5}}, and {\footnotesize\circled{6}}, respectively.  
One can easily observe that this trivial design is extremely inefficient and a straightforward improvement can be applied if we consider beam grouping based on adaptive beamforming~\cite{park:2013:comletter, biason:2017:multicast, biason:2019:multicast}. % or codebook ~\cite{naribole:2017:ton} methods. 
As shown in \fref{fig:hierarchical}, nodes that are closer to the source nodes (i.e., {\footnotesize\circled{1}}, {\footnotesize\circled{2}}, and {\footnotesize\circled{3}}) are served together with a wider beam, while the father nodes (i.e., {\footnotesize\circled{4}}, and {\footnotesize\circled{5}}) and the nodes that are not in proximity with the other nodes (i.e., {\footnotesize\circled{6}}) with narrower beams. Although adaptive method provides higher flexibility in grouping the receiving nodes, it however comes at the expense of more complex beamforming and costly antenna architecture. 

The above designs rely only on single-hop transmissions, which can be problematic in many practical scenarios. More specifically, there might exist nodes that are not reachable by the source or nodes that are not feasible for high transmission rates due to large distance (i.e., node {\footnotesize\circled{5}} in sector $4$) or the presence of blockages~\cite{lin:2015:multihoprelay, du:2017:backhaulrelay}. In such cases, a relay-aided transmission is inevitable to ensure reachability and guarantee high-performance multicasting (in terms of throughput and delay). With relay enabled, a node can serve as a transmitting node as soon as it receives the data from another node. As shown in~\fref{fig:relay}, upon receiving data from node {\footnotesize \circled{0}} in the first time slot, nodes {\footnotesize \circled{3}}, and {\footnotesize\circled{4}} act as the relay node for node {\footnotesize\circled{1} \circled{2}}, and {\footnotesize \circled{5}}, respectively. With this flexibility, we can break down a low-rate multicast transmission into a combination of multiple high-rate unicast and/or multicast transmissions that can be scheduled separately. Interestingly, we can then leverage the limited coverage of directional transmissions in mmWave due to the significantly increased spatial gain brought by significantly reduced interference among concurrent (unicast or multicast) transmissions; in \fref{fig:relay}, links $\footnotesize{{\circled{0}} {\tiny\rightarrow} {\circled{6}}}$, $\footnotesize{{\circled{3}} {\tiny\rightarrow}} {\footnotesize\circled{1}~\circled{2}}$, and $\footnotesize{{\circled{4}} {\tiny\rightarrow} {\circled{5}}}$ occur simultaneously. 

We believe the optimal performance of mmWave systems should jointly exploit all these properties of mmWave communication, namely \emph{relaying} and \emph{spatial sharing}. Thus far, the existing works have considered single aspects, but never jointly. This motivates us to design new mmWave multicast scheduling algorithms integrating both relaying and spatial sharing. Unsurprisingly, the joint optimization is complicated and the specific challenge resides in designing efficient communication group composition and spatial sharing scheduling. 
With both spatial and temporal factors involved, the relay nodes have to be determined gradually and the source and the relay nodes have to select carefully their target nodes depending on how the communication will affect the total completion time. This situation becomes even worse when only limited knowledge about the behavior of the other node with concurrent transmissions is available.  

To address these challenges, we provide a comprehensive model and an integer linear program (ILP) to characterize the problem, with the objective of minimizing the multicast completion time (i.e., the time required for all nodes to receive the intended data). The ILP aims to find the optimal scheduling policy that determines the transmitting nodes and their corresponding receivers at each time slot. Specifically, it jointly minimizes the duration of each time slot accounting for all \emph{concurrent transmissions}\footnote{In mmWave communication systems, the terminology of spatial sharing is also commonly referred to as concurrent transmission. In this paper, these terms are used interchangeably.} while selecting the optimal relay node. Exploiting spatial sharing in the relay-aided multicast transmission requires careful scheduling, both spatially and temporally, which is usually not of concern in the conventional multicast. Hence, the problem formulation for directional multicasting is significantly different and inherently more complicated than that of the conventional multicast scheduling in the literature. Ultimately, solving the ILP provides a tight lower bound for the multicast completion time in a mmWave network leveraging both relaying and spatial sharing gains.

To account for the deployment in real-world scenarios in equipment with computational power constraints and to ensure scalability, we further present a lightweight distributed algorithm, namely \texttt{mmDiMu}. The high-level idea is to exploit concurrency by allowing each transmitting node to autonomously decide and transmit to its target node(s), regardless of the other concurrent transmissions in the network. The set of target nodes for each transmitting node is determined based on the physical distance of nodes and is updated after every transmission time slot. 

The following summarizes the contributions of this paper:
\begin{itemize}
\item We identify the challenges and opportunities in mmWave multicast scheduling and provide an ILP formulation that finds the optimal scheduling policy by jointly leveraging relaying and spatial sharing gains.
\item Due to the exponential complexity of the ILP-based solution (namely \texttt{ILP}), we propose \texttt{mmDiMu} heuristic -- a \emph{scalable distributed mmWave multicast scheduling algorithm}. This lightweight algorithm has significantly lower complexity, and is more practical than \texttt{ILP}. 
\item We perform extensive simulations to validate the performance of our algorithm in both low- and high-density networks. As expected \texttt{ILP} demonstrates a substantial gain in completion time as compared to all other algorithms. While there is a slight gap between \texttt{mmDiMu} and \texttt{ILP} solutions, we can observe a significant improvement over the existing algorithms, i.e., \texttt{FHMOB} in~\cite{sim:2017:FHOMB}, and \texttt{OMS} in~\cite{low:2010:multicasting} for sub-6GHz and the adaptive beamwidth algorithm (i.e, \texttt{Adapt}) in~\cite{biason:2019:multicast} for mmWave, which to the best of our knowledge represents the state of the art. \item We evaluate interference imposed on unintended receivers by the proposed algorithm and show that the impact of interference is marginal even for high-density scenarios.
\item We also provide valuable insights on the design of a mmWave multicast system and design guideline depending on the network's density and system configurations.    
\end{itemize}

The rest of this paper is organized as follows. In \sref{sec:related_work}, we present the state of the art for multicast scheduling algorithms. \sref{sec:model} includes a description of the system model and its problem formulation. The optimal solution (i.e., based on ILP) is presented in \sref{sec:milp}, and \sref{sec:heuristic} presents a lightweight heuristic. The performance evaluation is presented in \sref{sec:result}. In \sref{sec:discussion}, we discuss other important aspects to design mmWave multicasting and \sref{sec:conclusion} concludes our paper.

\section{Related Work}
\label{sec:related_work}
As a key technology for beyond-5G networks, mmWave has been considered for many emerging applications (e.g., autonomous driving, public safety, and mobile video streaming) that typically require the distribution of data in large volume with low latency. Unfortunately, directional mmWave links suffer from limited coverage, and it complicates multicasting. Many existing works on mmWave mainly focus on unicast transmissions. With that said, the challenges and benefits of mmWave multicast remain understudied. In this section, we present the state of the art of multicast techniques for both sub-6GHz and mmWave networks, while differentiating them with our proposed approach.

\subsection{Sub-6GHz multicasting}
The most basic type of multicasting is broadcast, in which all nodes are served simultaneously. In this case, the transmit rate is limited by the node with the worst channel quality. Improving over this basic technique, many opportunistic multicast techniques are proposed in~\cite{kozat:2008:oppmul, tplow:2010:oms, sim:2016:tmc} and the references therein. 
These techniques exploit multiuser diversity by opportunistically transmitting to an arbitrary subset of the nodes with better instantaneous channel quality. 
As a result, they outperform the broadcast scheme and achieve higher throughput. 
However, this technique still suffers from poor performance when the network has nodes located at its edge.
In the extreme case (i.e., when many nodes are located at the edge), it performs similarly to a broadcast scheme. 

Overcoming the above issue, the research community has explored multicast beamforming. Multicast beamforming uses the beamforming technique that focuses the transmit signal power at only one direction of interest by adjusting the antenna gains. As a result, it improves the signal-to-noise ratio (SNR) of the nodes in that direction. 
Authors in~\cite{sen:2008:noclient} publish one of the first work on improving the system throughput with this technique. They first use omnidirectional multicast to transmit to nodes with better channel quality and then use directional multicast to transmit sequentially to the remaining nodes. 
To further improve the system performance, a better method applies beamforming weights at the antenna leading to the maximization of the worst SNR, at the expense of degrading the SNR of other nodes (i.e., the nodes that are located closer to the transmitter). Many research works demonstrate this technique yields a high system throughput \cite{low:2010:multicasting, chang:2008:maxminfair, aryafar:2013:adam} and minimizes completion time~\cite{sundaresan:2009:OBS, zhang:2012:switched, sim:2017:FHOMB}.    

The aforementioned works mainly focus on scheduling the subset of nodes in a system to achieve the intended goal, where neither coverage nor blockage is an issue. Specifically, a source node can simultaneously transmit to any arbitrary subset or even all nodes if desired. Nevertheless, operating at high frequency, mmWave communications are prone to extremely high attenuation and penetration loss. Furthermore, the use of directional transmission (which only covers a small angular area) makes it impossible to serve any arbitrary nodes in the system simultaneously. As a result, the multicasting techniques designed for sub-6GHz communication yield suboptimal performance for mmWave communication. To shed light on this aspect, we specifically benchmarked the performance of our proposed algorithms to two seminal multicast schedulers used in sub-6GHz systems (i.e., in~\cite{low:2010:multicasting, sim:2017:FHOMB}) in \sref{sec:result}.  
 
\subsection{mmWave multicasting}
An initial work addressing the need for the redesign of mmWave multicast scheduling is presented in \cite{park:2013:comletter} where the authors emphasize on the use of adaptive beamwidth to improve the grouping of the multicast nodes to achieve higher throughput. Similar work is presented in \cite{biason:2019:multicast} where the authors investigate the trade-off between transmission beamwidth and achievable SNR to ensure high throughput. These schedulers may require a high level of beamwidth adaptation to form arbitrary beams to provide coverage to the multicast nodes. Therefore, this design increases the complexity and the cost of the antenna design. In contrast, with a highly reduced complexity, the authors in~\cite{naribole:2017:ton} present a practical IEEE 802.11ad compliance approach where a codebook-based scheduler with one radio frequency (RF) chain is applied.      

All above-mentioned works consider only single-hop multicasting in which the multicast transmission rate remains limited to the nodes located farthest from the source node without leveraging spatial sharing.      
Later, the benefits of relay and spatial sharing are separately considered in~\cite{chu:2016:sis} and \cite{feng:2017:iwcmc} to improve the multicast rate and spectral efficiency, respectively. In~\cite{chu:2016:sis}, the authors exploit relaying only to overcome non-line-of-sight paths, but not for performance optimization. In~\cite{feng:2017:iwcmc}, the authors leverage spatial sharing in which they enable the simultaneous transmission of single-hop unicast and multicast sessions to increase network efficiency. 

To sum up, all the works mentioned above works either consider \emph{multi-hop relay} or \emph{optimal spatial sharing}, but not jointly. \emph{To the best of our knowledge, we are the first to jointly consider both to minimize the data delivery time for mmWave multicast communications}.

\section{System Model and Assumptions}
\label{sec:model}

We consider a mmWave network composed of $N+1$ randomly distributed nodes denoted by set $\mathcal{N}= \{0, 1,..., N \}$, 
where node $0$ represents the source and the other nodes $n = 1,...,N$ are interested in receiving data of size $B$ from the source. We assume relaying is enabled in the network, meaning that all the nodes, once receiving the data, can transmit the data to other nodes. We consider a time slotted system where the number of time slots for multicasting the data is denoted by variable $S$, and the set of time slots is given by $\mathcal{S} = \{1,...,S\}$. The length of each time slot is not necessarily equal, but we ensure that transmissions happen only within one-hop at each time slot. To exploit spatial sharing, multiple concurrent transmissions can coexist at each time slot.

We call a node that transmits data to other node(s) a parent node (PN), and we denote by $\mathcal{P}^s \subset \mathcal{N}$ the set of PNs at time slot $s$. Inversely, a node that receives data is called a child node (CN), and we denote by $\mathcal{C}_m^s \subset \mathcal{N}$ the set of CNs of PN $m$ at time slot $s$. A node can serve as PN in multiple time slots, and the data has to be completely delivered to all its CNs in each of these time slots. Therefore, we have $\mathcal{C}_m^{s_1} \neq \mathcal{C}_m^{s_2}$ for any $s_1, s_2 \in \mathcal{S}$ and $s_1 \neq s_2$. Each node in the network has a \emph{fixed} transmit power and $L$ equal-width orthogonal lobes numbered counterclockwise starting from $0^{\circ}$, denoted by $\mathcal{L} = \{1,...,L\}$. For each node $m \in \mathcal{N}$, we denote by $\mathcal{N}_m^l$ the set of nodes that are within the coverage of lobe $l \in \mathcal{L}$ of the node. For example, we have $\mathcal{N}_0^1 = \{1,2\}$ and $\mathcal{N}_3^4 = \{5\}$ in \fref{fig:sample_net}. Note that, as the lobes are orthogonal, a node can activate more than one lobe simultaneously.

\begin{figure}[!t]
	\centering
	\begin{tikzpicture}
		\filldraw[fill=gcolor0, draw=gray!40] (0,0) -- (40mm,0mm) arc (0:45:40mm) -- cycle;
		\filldraw[fill=gcolor0, draw=gray!40] (0,0) -- (28.28mm,28.28mm) arc (45:90:40mm) -- cycle;
		\filldraw[fill=gcolor0, draw=gray!40] (0,0) -- (0mm,40mm) arc (90:135:40mm) -- cycle;
		\filldraw[fill=gcolor0, draw=gray!40] (0,0) -- (-28.28mm,28.28mm) arc (135:180:40mm) -- cycle;
		\filldraw[fill=none, draw=gray!20] (-0.5,1.5) -- (15mm,15mm) arc (0:45:20mm) -- cycle;
		\filldraw[fill=none, draw=gray!20] (-0.5,1.5) -- (9.14mm,29.14mm) arc (45:90:20mm) -- cycle;
		\filldraw[fill=none, draw=gray!20] (-0.5,1.5) -- (-5.0mm,35mm) arc (90:135:20mm) -- cycle;
		\filldraw[fill=none, draw=gray!20] (-0.5,1.5) -- (-19.14mm,29.14mm) arc (135:180:20mm) -- cycle;
		\filldraw[fill=gcolor1, draw=gray!20] (0,0) -- (0mm,18.81mm) arc (90:135:18.81mm) -- cycle; 	% sector 5
		\filldraw[fill=gcolor2, draw=gray!20] (0,0) -- (28.5mm, 0mm) arc (0:45:28.5mm) -- cycle; 		% sector 3
		\filldraw[fill=gcolor3, draw=gray!20] (-0.5,1.5) -- (-19.14mm, 29.14mm) arc (135:180:20mm) -- cycle; 	% sector 3
		\filldraw[fill=gcolor5, draw=gray!20] (-0.5,1.5) -- (10mm, 15mm) arc (0:45:15mm) -- cycle; 		% sector 3
		\draw[gcolor5, thick, dotted] (0,0) --  node[circle, inner sep=0pt, fill=gcolor1, minimum size=3mm, font=\footnotesize] {$1$} (-0.5,1.5);
		\draw[gcolor5, thick, dotted] (0,0) --  node[circle, inner sep=0pt, fill=gcolor2, minimum size=3mm, font=\footnotesize] {$3$} (2.3,1.1);
		\draw[gcolor0, thick, dotted] (-0.5,1.5) --  node[circle, inner sep=0pt, fill=gcolor3, minimum size=3mm, font=\footnotesize] {$2$} (-2,2.4);
		\draw[gcolor0, thick, dotted] (-0.5,1.5) --  node[circle, inner sep=0pt, fill=gcolor5, minimum size=3mm, font=\footnotesize] {$1$} (0.6,2.1);
		\node at (0, 0) [circle, inner sep=0pt, minimum size=4mm, draw=gray!50, fill=white, font=\small]{0};
		\node at (1.8, 0.5) [circle, inner sep=0pt, minimum size=4mm, draw=gray!50, fill=white, font=\small]{1};
		\node at (2.3, 1.1) [circle, inner sep=0pt, minimum size=4mm, draw=gray!50, fill=white, font=\small]{2}; 
		\node at (-0.5, 1.5) [circle, inner sep=0pt, minimum size=4mm, draw=gray!50, fill=white, font=\small]{3};
		\node at (0.6, 2.1) [circle, inner sep=0pt, minimum size=4mm, draw=gray!50, fill=white, font=\small]{4};
		\node at (-2, 2.4) [circle, inner sep=0pt, minimum size=4mm, draw=gray!50, fill=white, font=\small]{5};
		\node at (3.6, 1.5) [circle, inner sep=0pt, minimum size=2.5mm, draw=gray!0, fill=white, font=\footnotesize, text=gcolor5]{$l$=$1$};
		\node at (1.49, 3.6) [circle, inner sep=0pt, minimum size=2.5mm, draw=gray!0, fill=white, font=\footnotesize, text=gcolor5]{$l$=$2$};
		\node at (-1.49, 3.6) [circle, inner sep=0pt, minimum size=2.5mm, draw=gray!0, fill=white, font=\footnotesize, text=gcolor5]{$l$=$3$};
		\node at (-3.6, 1.5) [circle, inner sep=0pt, minimum size=2.5mm, draw=gray!0, fill=white, font=\footnotesize, text=gcolor5]{$l$=$4$};
		
	\end{tikzpicture}
	\vspace{-2mm}
 	\caption{Data dissemination via multicast scheduling for $L=8$. The number on the edges indicates the number of required transmission slots.}
	\label{fig:sample_net}
	\vspace{-8mm}
\end{figure}
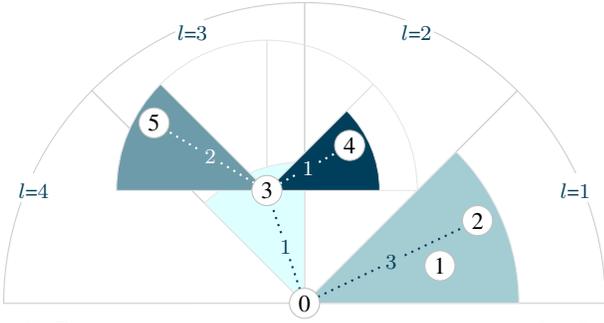

We adopt a path-loss model used in~\cite{maccartney:2017:globecom} (will be detailed in Section~\ref{sec:sim_setup}), 
and the received rate is computed using the Shannon capacity model from~\cite{Akdeniz:2014:jsac}.
We denote by $\gamma_{m,n}$ the SNR of the signal received at CN $n$, transmitted from PN $m$.
A node is called a target node (TN) of a PN if its received signal has the lowest SNR as compared to the other CNs within the same lobe of the PN.
In fact, the nodes with SNR worse than that of the TN are assumed to be unable to decode the message transmitted by the PN. 
Note that there is at most one TN in each lobe for a PN.
We denote by $\mathcal{G}_m^s$ the set of all TNs of PN $m$ at time slot $s$.
Given $\mathcal{C}_m^s$, the set $\mathcal{G}_m^s$ can be formally defined as
\begin{equation}
\label{eq:targetSet}
\mathcal{G}_m^s =  \{ n~\vert~   \gamma_{m,n} = \min_{ u }   \{ \gamma_{m, u} \}, u \in \mathcal{N}_m^l \cap \mathcal{C}_m^s, \forall l \in \mathcal{L} \}.
\end{equation}
Note that $|\mathcal{G}_m^s| = 0$ means that node $m$ does not transmit at time slot $s$ 
and $|\mathcal{G}_m^s|= L$ means that node $n$ steers its beam towards all directions, where $|\cdot|$ gives the cardinality of a set.
Since the TNs experience the worst channel conditions in comparison to other CNs within the same lobe, the maximum rate which determines the transmission time of a PN depends on the SNR of the set of its TNs $\mathcal{G}_m^s \subseteq \mathcal{C}_m^s$.
Given the TN set $\mathcal{G}_m^s$ of a PN $m$, finding the optimal transmitting rate for the PN is as discussed in \cite{sim:2017:FHOMB}. 
Our focus is on obtaining the optimal $\mathcal{C}_m^s$ for each node $m$ at each time slot $s$.
Note that activating more lobes simultaneously results in lower transmission rate.
Let $r^*(\mathcal{G}_m^s)$ be the optimal transmit rate. The time required for PN $m$ to complete the data transmissions to all its TNs (including its CNs) at time slot $s$ is given by
\begin{equation}
\label{eq:time_toTN}
t_m(\mathcal{G}_m^s) = \frac{B}{r^*(\mathcal{G}_m^s)}.
\end{equation}
At each time slot, multiple PN can transmit simultaneously, exploiting spatial sharing.
As a result, the duration of a time slot is determined by the longest transmission at the time slot, that is,
\begin{equation}
 \label{eq:time slot-s}
t^{s}(\mathcal{N}) = \max_{m \in \mathcal{N}} \quad \left\{  t_m (\mathcal{G}_m^s) \right\}.
\end{equation}
Our objective in this work is to minimize the total duration of all the time slots in $\mathcal{S}$, namely multicast \emph{completion time}, 
by jointly minimizing $t^s$ and $S$, and to determine the set of PNs and their corresponding CNs in each time slot. 
The completion time $T$ can be expressed by 
\begin{equation}
	\label{eq:completion_time}
T(\mathcal{N}) = \sum_{s \in \mathcal{S}} t^s(\mathcal{N}).
\end{equation}
The following constraints should be considered. % respected.
First, all nodes have to receive the data within $S$ time slots, i.e., 
\begin{equation}
\underset{m \in \mathcal{N}, s \in \mathcal{S}}{\bigcup} \mathcal{C}_m^s = \mathcal{N}\setminus \{0\}.
\end{equation}
Then, a node can only transmit data to other nodes if it has already received the data, i.e., 
\begin{equation}
\forall s \geq 2, m \in \mathcal{P}^s~\implies~m \in \underset{\substack{x \in \mathcal{P}^{s'}\\ 1 \leq s' \leq s-1}}{\bigcup} \mathcal{C}_x^{s'} \cup \{0\}.
\end{equation}

\section{Proposed Approaches}

In this section, we describe our solutions to the min-time mmWave multicast scheduling problem. 
We first provide an ILP formulation that gives an optimal schedule, and then we propose a more scalable distributed algorithm.

\subsection{Optimum Solution by ILP}
\label{sec:milp}
We first define terms and variables using a toy example in \fref{fig:sample_net}. We define $K$ as the number of elements in the power set of $\mathcal{N}\setminus \{0\}$, excluding the empty set, i.e., $K = 2^N-1$. In \fref{fig:sample_net}, we have $N = 5$, $K = 31$, and $L = 8$.

\begin{itemize} 
\item $\mathbf{g}_m^s$ (target vector of PN $m$ in time slot $s$):
a binary vector  $\mathbf{g}_m^s = [g_{m,1}^s  \dots  g_{m,N}^s]^\intercal \in \{0,1\}^N$ in which  $(.)^\intercal$ is the transpose operator and  $g_{m,n}^s = 1$ if node $n$ is a TN of PN $m$ in time slot $s$.
For example, in Fig. \ref{fig:sample_net}, nodes {\scriptsize\circled{2}} and {\scriptsize\circled{3}} are  the TNs of the source in the first time slot, and hence the target vector is $\mathbf{g}_\mathrm{0}^1 = [0 1 1 0 0]^\intercal$.
There are $K$ possible combinations for a target vector for each PN. %if a node acts as a PN.

\item $\mathbf{U}$ (target matrix): a binary matrix of size $N \times K$. 
Each of the columns of $\mathbf{U}$ represents a possible choice for a target vector, where $\mathbf{g}_m^s$ is a column of $\mathbf{U}$. In fact, $\mathbf{U}$ is independent of the nodes, and it shows the state-space of the target vector $\mathbf{g}_m^s, m\in \mathcal{N}$. Precisely, $\mathbf{U} = [\mathbf{u}_1^\intercal, \dots, \mathbf{u}_K^\intercal]$ where $\mathbf{u}_k$ is a $1 \times N$ binary vector. We form $\mathbf{U}$ by filling $\mathbf{u}_k$, $1 \leq k \leq K$, via the reverse ({\small \texttt{rev}}) of the $N$-bit  decimal-to-binary ({\small\texttt{dec2bin}}) conversion of the index $k$.
For instance, $\mathbf{u}_6 = {\small \texttt{rev} } ([ {\small \texttt{dec2bin}(6)}])= {\small \texttt{rev}} ([0 0 1 1 0])= [01100]$ and in \fref{fig:sample_net}, based on the definition of TN in \eqref{eq:target}, we have $\mathbf{g}_\mathrm{0}^1 = \mathbf{u}_6^\intercal$. 

\item $\mathbf{p}_m^s$ (PN vector of PN $m$ in time slot $s$): a binary vector $\mathbf{p}_m^s = [p_{m,1}^s, \dots, p_{m,K}^s]^\intercal \in \{0,1\}^K, \forall m \in \mathcal{N}$  and $||\mathbf{p}_m^s|| \leq 1$. If node $m$ is a PN at time slot $s$, then $||\mathbf{p}_m^s|| = 1$, otherwise, $||\mathbf{p}_m^s||= 0$. Precisely, $p_{m,k}^s =1$ if PN $m$ chooses the $k$-th column of $\mathbf{U}$ as its target vector. Given $\mathbf{p}_m^s$, the TNs of PN $m$  is obtained by
\begin{equation}
\label{eq:target}
\mathbf{g}_m^s  =  \mathbf{U}  \mathbf{p}_m^s. 
\end{equation}

\item $\mathbf{N}_m$ (observation matrix): $\mathbf{N}_m = [\mathbf{n}_m^1,..., \mathbf{n}_m^L]$ is a binary matrix of size  $N \times L$, defined for every $m \in \mathcal{N}$. For each node $m$, $\mathbf{N}_m$ indicates with which lobe can node $m$ cover the other nodes using a single-hop transmission. Precisely, $\mathbf{N}_m(n,l) = 1$ if node $n$ is within lobe $l$ of node $m$. For network in \fref{fig:sample_net}, we have 
\begin{equation} \label{eq:os}
\mathbf{N}_\mathrm{0} = 
\footnotesize
\begin{bmatrix}
1&	0&	0&	0&	0&	0&	0&	0\\
1&	0&	0&	0&	0&	0&	0&	0\\
0&	0&	1&	0&	0&	0&	0&	0\\
0&	1&	0&	0&	0&	0&	0&	0\\
0&	0&	1&	0&	0&	0&	0&	0\\
\end{bmatrix}.
\end{equation}

\item $\mathbf{C}_m$ (CNs matrix):
a binary matrix of size $N \times K$ that shows if a PN $m$ transmits to its TNs, which of the other nodes fall within the coverage area of the PN.
While $\mathbf{g}_m^s$ represents the set $\mathcal{G}_m^s$ of TNs of PN $m$, defined in \eqref{eq:target}, $\mathbf{C}_m$ corresponds to the set $\mathcal{C}_m^s$ of CNs of PN $m$, which can also be served given the TNs in $\mathbf{g}_m^s$. 
Let $\mathbf{g}_m^s$ be the target vector of PN $m$ corresponding to the $k$-th column of $\mathbf{U}$, then the elements of the $k$-th column of $\mathbf{C}_m$, which are equal to 1, represent all the nodes which can be served by such a target vector.
To clarify, let node $n \in \mathcal{N}_m^l$ be a target node of PN $m$ given $\mathbf{g}_m^s$ corresponding to the $k$-th column of $\mathbf{U}$.
Based on the definition, since node $n$ as the TN of PN $m$ is always in $\mathcal{C}_m^s$, then,  $\mathbf{C}_m(n,k) = 1$.
Further, we have $\mathbf{C}_m(u,k) = 1$ if  $\gamma_{m,u} \geq \gamma_{m,n}$, $\forall u \in \mathcal{N}_m^l$.
Based on item (ii), for the source node in Fig. \ref{fig:sample_net}, we have $\mathbf{C}_0(:,6) = [11100]^\intercal$ which corresponds to $\mathbf{u}_6^\intercal$. %(if we you exceed the page limit, remove this sentence.)} 
Given the PN vector $\mathbf{p}_m^s$, we denote all the CNs, covered by PN $m$, by a binary vector  $\hat{\mathbf{c}}_m^s = [\hat{c}^s_{m,1}, \dots, \hat{c}^s_{m,N}]^\intercal$ where $\hat{c}^s_{m,n} = 1$ if node $n$ is covered by PN $m$ at time slot $s$.  $\hat{\mathbf{c}}_m^s$ is thus obtained by
\begin{equation}
\label{eq:cvrd}
\hat{\mathbf{c}}_m^s = \mathbf{C}_m \mathbf{p}_m^s.
\end{equation}
 
\item $\mathbf{t}_m $ (transmission duration): $\mathbf{t}_m = [t_{m,1},..., t_{m,K}] \in  \mathbb{R}^K, m \in \mathcal{N}$, a real-valued vector . 
If a PN $m$  chooses the $k$-th column of $\mathbf{U}$ as its target vector  $\mathbf{g}_m^s$, then, $t_{m,k}$ shows the duration of transmission defined in \eqref{eq:time_toTN}.
\end{itemize}

Matrices $\mathbf{U}, \mathbf{N}_m,  \mathbf{C}_m, \mathbf{t}_m$ can be calculated given the distribution of nodes in the network,
while $\mathbf{p}_m^s, \forall m \in \mathcal{N}, s \in \mathcal{S}$ are to be found by the ILP.
Using these terms, the ILP formulation is provided as follows.
\begin{subequations}\label{eq:ILP}
\begin{mathalign}
& \underset{ p_{m,k}^s } {\text{min}}
& & T(\mathcal{N}) = \sum_{ s \in \mathcal{S}} \max_{ m \in \mathcal{N}}  \{   \mathbf{t}_m \mathbf{p}_m^s \} \\
& \text{s.\,t.} & &  
\sum_{k=1}^{K} p_{m,k}^s= 
\begin{cases}
1 & m = 0, s = 1  \\
0 & m = [1,...,N], s = 1 \\
\leq 1 & s \geq 2
\end{cases}
\label{eq:const_a}  \\
& & &  \sum_{k=1}^{K} p_{m,k}^s  \! \leq \! \! \sum_{s^\prime = 1}^{s-1} \sum_{{x \in \mathcal{N}\backslash \{m\}}} \hat{c}_{x,m}^{s^\prime},  \forall m \in \mathcal{N} \! \! \setminus \! \!  \{0\}, s \geq 2 \label{eq:mustbecovered} \\ 
& & & (\mathbf{g}_m^s)^\intercal \mathbf{n}_m^l \leq 1, \qquad \qquad \forall m \in \mathcal{N}, \forall s, \forall l \label{eq:const_b2} \\
& & & p_{m,k}^s \in \{0,1\}, \qquad \qquad \forall m \in \mathcal{N},  \forall s, \forall k \label{eq:const_last} 
\end{mathalign}
\end{subequations}

As mentioned, $p_{m,k}^s \in \{0,1\}$ in \eqref{eq:ILP} is the decision variable, which determines the TNs of PN $m$ as in~\eqref{eq:target}.
\eqref{eq:const_a} expresses that the source node must transmits at $s = 1$, but not the other nodes.
In the following time slots, any of the nodes in $\mathcal{N}$ could be a PN given that it has received the data in any previous time slots $1\leq s^\prime \leq s-1$; the constraint in~\eqref{eq:mustbecovered} indicates this.
Finally,  \eqref{eq:const_b2} guarantees that the number of TN in a lobe is at most one.

Regarding the complexity, ILP formulation is an NP-hard problem as a special case of the problem has been shown to be NP-hard \cite{Zagalj:2002:MobiCom}. Although NP-hard, its running time depends on the number of integer variables. 
Our proposed ILP has $N\times(2^N+1)$ variables, and thus a complexity of $O(2^N)$, which exponentially increases with $N$. 
Clearly, the ILP-based solution has an exponential time complexity, and it can only be solved for very small problem instances (i.e., small $N$). 
For this reason, in the next section, we design a practical and lower complexity heuristic.

\subsection{Distributed Multicast Scheduling}
\label{sec:heuristic}

Our distributed multicast scheduling heuristic, namely \texttt{mmDiMu}, accounts for both relay and spatial sharing. 
By having each PN deciding autonomously its CNs to transmit to, \texttt{mmDiMu} is scalable and distributed in nature as opposed to the centralized ILP solution. The pseudocode of the algorithm is as shown in \aref{algo:mmDeMu}. In what follows, we elaborate on the detail of the algorithm. 

We use $\mathcal{W}$ to denote the set of waiting nodes that have not received the intended data.
Initially, i.e., at the first time slot, node $0$ is the only PN in set $\mathcal{P}^1$, 
and we have $\mathcal{W} = \{1,...,N\}$. 
We use $\mathbf{D}$ to denote the distance matrix, where $\mathbf{D}(m,n)$ represents the distance between nodes $m$ and $n$.
At each of the following time slots $s \geq 2$, we select for each node $n \in \mathcal{W}$ the PN $m$ in $\mathcal{P}^s$ with the least distance $\mathbf{D}(m,n)$.
In the case where a node is equidistance from two or more PNs, it will randomly select one of the PNs.
After this process, for each node $m \in \mathcal{P}^s$ we obtain its CN set $\mathcal{C}_m^s$ at this time slot, 
and we apply the opportunistic multicast scheduling that maximizes the sum throughput to select the set of nodes from $\mathcal{C}_m^{s}$ for PN $m$ to transmit to. The intuition lies in maximizing the achievable rate for each transmission session to promote minimum session transmission time, and thus resulting in minimum completion time.
Once receiving the data, a node will be removed from the set $\mathcal{W}$ and added to the PN set $\mathcal{P}^{s+1}$.
The above process is repeated until all nodes receive the data.
In each time slot, the time for each transmission is recorded as $t_m(\mathcal{C}_m^{s*})$, where $\mathcal{C}_m^{s*}$ is the optimal. % based on \cite{tplow:2010:oms}.  
The multicast completion time thus can be calculated as $\sum_{s \in \mathcal{S}} \max_{m \in \mathcal{N}} \left\{  t_m(\mathcal{C}_m^{s*}) \right\}$.

\begin{algorithm}[t]
\begin{algorithmic}[1]
\footnotesize
\State Input:  $N$, $\mathbf{D}$
\State Initialize counters: time slot $ s\leftarrow1 $ 
\State Initialize waiting node set: $ \mathcal{W} \leftarrow \{1,...,N\} $
\State Initialize PN sets: $ \mathcal{P}^s \leftarrow \{0\}, \forall s $ 
\State Initialize CN sets: $ \mathcal{C}_m^s \leftarrow \emptyset, \forall m, \forall s $ 

\While { $ \mathcal{W} \neq \emptyset $ }
	\ForAll{ $m \in \mathcal{W}$ }
    \State {Select PN: $  m = \arg\min_{u \in \mathcal{P}} \mathbf{D}(u,n) $} 
	\State {Store CN: $\mathcal{C}_m^s \leftarrow  \mathcal{C}_m^s \cup n $ }
	\EndFor
  
  	\ForAll{ $m \in \mathcal{P}^{s}$ }
    \State Select CN set with max-throughput: $\mathcal{C}_m^{s*}$ is served with rate $r^s_m$  
	\State Update waiting set: $\mathcal{W} \leftarrow \mathcal{W} \backslash \mathcal{C}_m^{s*} $
	\State Update PN set: $\mathcal{P}^{s+1} \leftarrow \mathcal{P}^s \cup \mathcal{C}_m^{s*} $
	\State Compute transmission time of PN $m$: $t_m(\mathcal{C}_m^{s*}) \leftarrow B / r^s_m $
  	\EndFor
\State $s \leftarrow s + 1$
\State {Compute slot-time: $t^s = \max_{m \in \mathcal{P}^{s}} t_m(\mathcal{C}_m^{s*}) $ }
\EndWhile
\State \textbf{end}
\State {Output multicast completion time $T = \sum_s^S t^s$ }
\end{algorithmic}
\caption{Pseudocode of \texttt{mmDiMu} algorithm}
\label{algo:mmDeMu}
\end{algorithm}

\section{Performance Evaluation}
\label{sec:result}

This section evaluates the performance comparisons between the baseline and our proposed multicast algorithms.  

% ===========================================================================
%	Simulation setup
% ===========================================================================
\subsection{Simulation Setup}
\label{sec:sim_setup}
We consider a uniform and randomly distributed nodes within a $200$m$\times200$m area with the source node (i.e., PN $0$) located at the center. 
We adopt the mmWave path-loss model in~\cite{maccartney:2017:globecom}, which is written as,
\begin{equation}
\text{PL} [\text{dB}] = \alpha + 10\beta\log_{10}(d_{m,n}) + 20\log_{10} (f_c) + \chi_{\sigma} ,  
\end{equation}
where $d_{m,n}$ is the distance between the PN $m$ and CN $n$, $f_c$ is the carrier frequency, and $\chi_{\sigma}$ represents the shadow fading with zero-mean Gaussian random variable and standard deviation $\sigma$ in dB. 
The received rate is computed using the Shannon capacity model in~\cite{Akdeniz:2014:jsac}.
\tref{tab:param} summarizes the parameter values used in the simulator. 

\begin{table}[!t]
\tiny
\caption{Channel parameters.}
\vspace{-2mm}
\centering
\begin{tabular}{|l|c|}
\hline
{\bf Parameter}					& {\bf Value} 								\\ \hline\hline
Free space path loss ($\alpha$)	& $32.4 \text{dB}$							\\ \hline
Carrier frequency ($f_c$)		& $ 73$GHz 									\\ \hline
System Bandwidth ($W$)			& $ 1 $GHz									\\ \hline
Transmit power					& $ 14.9$dBm \cite{maccartney:2017:jsac}	\\ \hline
Noise figure					& $ 4 \text{dB}@\text{PN}, 7 \text{dB}$@CN	\\ \hline
Thermal noise					& $ -174 \text{dBm/Hz} $					\\ \hline
Path loss exponent ($\beta$)	& $ 2.0$									\\ \hline
Standard deviation ($\sigma$)	& $ 1.9 \text{dB}$							\\ \hline
\multirow { 3}{*} {Shannon capacity ($\rho$)}		& $\rho = W \times \min \{ \log_2 \left( 1 + 10^{0.1(\text{SNR}-\delta)}\right), \rho_{\text{max}}\}$ \\
& maximum spectral efficiency $\rho_{\text{max}} = 4.6 \text{bps/Hz}$ \\
& loss factor $\delta = 1.6\text{dB}$ \\ \hline
Frame size ($B$)					& $1 \text{Gbits}$		\\ \hline
\end{tabular}
\label{tab:param}
\vspace{-5mm}
\end{table} 

% ===========================================================================
%	Benchmark algorithm
% ===========================================================================
\subsection{Benchmarked Algorithms}
This subsection highlights the different algorithms used in the performance comparison.

\noindent \texttt{ILP}.
This is based on solving the ILP presented in \sref{sec:milp}. It selects the transmission at each time slot, which globally maximizes the spatial sharing gain while achieving minimum completion time $T$. Therefore, it provides the lower bound for $T$.  
We solve the ILP  by employing Gurobi\footnote{http://www.gurobi.com/} along with CVX\footnote{http://cvxr.com/} in MATLAB  environment.
\\
\noindent \texttt{mmDiMu}.
This is our distributed algorithm that considers both relaying and spatial sharing. While suboptimal, \texttt{mmDiMu} scales well regardless of the network density. The detail of the algorithm is as presented in \sref{sec:heuristic}. 
Unlike \texttt{ILP}, \texttt{mmDiMu} uses a distributed approach, in which each PN makes the transmission decision autonomously.
\\
\noindent \texttt{OMS}~\textnormal{\cite{low:2010:multicasting}}.
This algorithm is a sub-category of a multicast with adaptive beamwidth scheduling algorithm. It provides optimal performance for multicast applications in conventional networks, capitalizing on the opportunistic gain. Essentially, \texttt{OMS} sorts the nodes according to their channel SNR and serves the subset of nodes that maximizes the instantaneous sum throughput.  
\\
\noindent \texttt{FHOMB}~\textnormal{\cite{sim:2017:FHOMB}}.     
Finite horizon opportunistic multicast beamforming (\texttt{FHOMB}) is designed specifically to minimize the completion time when sending a finite number of packets to multicast receivers. At each time slot, a subset of nodes is selected such that the estimated completion time is minimized. The estimated completion time is obtained by maximizing the minimum rate using multi-lobe beam; this beam multicasts (usually at a low broadcast rate) to the remaining receivers. 
\\
\noindent \texttt{Adapt}~\textnormal{\cite{biason:2019:multicast}}. 
This is a scalable heuristic which groups the multicast nodes in subgroups using a hierarchical structure to construct the multicast tree. An example scheduling is as depicted in \fref{fig:hierarchical}. Once the subgroups/beam are determined, the source node serves each multicast subgroup sequentially through the beams; the transmit rate at each beam is thus limited by the node with the lowest SNR within each beam.

\subsection{Evaluation Settings}
To evaluate the performance of each algorithm, we examine the impact of two main parameters: (1) the number of nodes $N$ and (2) the beamwidth $w = 360^{\circ}/L$ at the transceivers.
Due to the high complexity of ILP, i.e., $O(2^N)$, $N$ is restricted to $10$ in scenarios where ILP is involved for comparison. The rest of the algorithms are evaluated for up to $N = 100$. 
We evaluate the performance for transmitter beamwidth $w = \{15^{\circ}, 30^{\circ},45^{\circ}, 60^{\circ}, 90^{\circ}\}$. Note that, the transmit beamwidth $w$ has an impact on the transmission gain~\cite{antenna:gain}, which we account for in the computation of the receiving rate.
Unless mentioned otherwise, at the receiver side, we assume that it uses a quasi-omnidirectional mode for receiving.  

To ensure fair performance comparison between the algorithms, all algorithms use the same simulation setting. The minimum beamwidth is determined by $w$ in each simulation scenario in \sref{sec:sim_result}, and the beamwidth resolution is thus multiple of $w$ for all the algorithms except \texttt{Adapt}. Since \texttt{Adapt} operates based on adapting its beamwidth to the multicast group, it can freely adjust its beamwidth as long as the minimum beamwidth is $w$. For instance, when the simulation has a setting of $w = 45^{\circ}$, \texttt{Adapt} could have any beamwidths between $45^{\circ}$ and $360^{\circ}$ while the other algorithms could only have beamwidths that are a multiple of $45^{\circ}$, i.e., $\{90^{\circ}, 135^{\circ}, ... , 360^{\circ}\}$.

We implemented all the algorithms in Matlab and conducted the comparisons using the above settings. For each data point, we average the data over $200$ simulation runs and compute the corresponding $95\%$ confidence interval.

\subsection{Simulation Results} 
\label{sec:sim_result}
As defined in \eqref{eq:completion_time} in~\sref{sec:model}, the completion time $T$ is the time required for all network nodes to finally receive the multicast data (by summing up the duration $t^{\mathrm{s}}$ at all time slots). Specifically, it is represented by the time, at which the last multicast node receives its data. 

% ===========================================================================
%	Impact of increasing the number of users
% ===========================================================================
\subsubsection{Impact of the number of nodes $N$} 
Here, we evaluate the impact of different $N$, $N=\{ 2, 4, 6, 8, 10\}$, on the completion time $T$ by fixing the transceivers beamwidth $w = 45^{\circ}$.  

As a general trend, \fref{fig:var_n} shows that increasing the number of nodes $N$ also increases the completion time $T$. When $N$ is large, the number of multicast slots required to transmit to all the nodes increases as well. \texttt{ILP} performs best as it picks the best policy which results in minimum $T$, as formulated in \eqref{eq:completion_time}. It indeed only requires $23.54\%$, $3.79\%$, and $30.33\%$ of the multicast completion time required by \texttt{OMS}, \texttt{FHOMB}, and \texttt{Adapt}, respectively, for $N = 10$. Specifically, \texttt{ILP} achieves a reduction in completion time by up to $96.21\%$ as compared to the other algorithms. Our proposed algorithm \texttt{mmDiMu} also demonstrates a high gain in completion time. It achieves completion time reduction of up to $66.78\%$, $94.65\%$, and $62.77\%$ over \texttt{OMS}, \texttt{FHOMB}, and \texttt{Adapt}, respectively.  

Interestingly, while \texttt{OMS} performs well in conventional single-hop systems, it performs slightly worse than \texttt{Adapt} as $N$ increases. As $N$ increase, so as the SNR diversity of the nodes. In such a case, \texttt{OMS} will first opportunistically transmit to the node that has higher SNR. This behavior results in excluding the nodes with low SNR initially. As a result, it suffers from low transmitting rate at a later time; it still has to serve the remaining nodes that have lower SNR. Unlike \texttt{OMS}, \texttt{Adapt} groups the nodes based on angular and then divides the group to minimize the transmission time and form a binary tree structure. Therefore, it refrains from the suboptimality that comes from greedily scheduling the nodes with better SNR. On the other hand, \texttt{OMS} performs better than \texttt{Adapt} for smaller $N$ because the probability of having nodes at the edge is much smaller. Furthermore, \texttt{OMS} may use more than one (disjoint) beam to serve all the nodes, while this option is unavailable in \texttt{Adapt}. Therefore, sparse distribution of nodes -- this mostly occur when the node density is low (i.e., small $N$) -- harms the performance of \texttt{Adapt}.

\begin{figure}[t!]
	\centering
    \includegraphics[width=0.8\columnwidth]{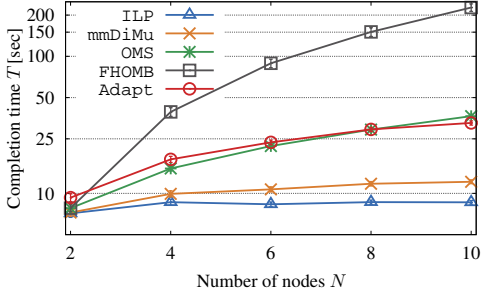}
    \vspace{-2mm}
	\caption{Completion time $T$ for different $N$ with $w = 45^{\circ}$.}
	\label{fig:var_n}
	\vspace{-7mm}
\end{figure}

Similarly, \texttt{FHOMB} in~\cite{sim:2017:FHOMB} that performs well in single-hop multicasting, performs poorly here. In \texttt{FHOMB}, a node receives the complete frame over multiple fixed-length time slots. %(as demonstrated in Fig. 1.b). 
At each slot, the policy (i.e., the subset of nodes to transmit to) which gives the lowest estimated completion time (up to the time all nodes received the frame) is chosen. As mentioned, to determine the estimated completion time, the remaining nodes are served with broadcast. In mmWave networks, broadcasting in all direction results in a very low transmission rate. 
Therefore, the estimated completion time is significantly longer than a slot time. Here, lower estimated time is favored since it provides a lower total transmission time. In most cases, this comes at the expense of a long slot duration $t^s$.  
As seen in \fref{fig:var_n}, this results in high completion time. 

\begin{figure}[t!]
\centering
	 \begin{subfigure}{0.23\textwidth}
    	\centering
    	\includegraphics[width=\columnwidth]{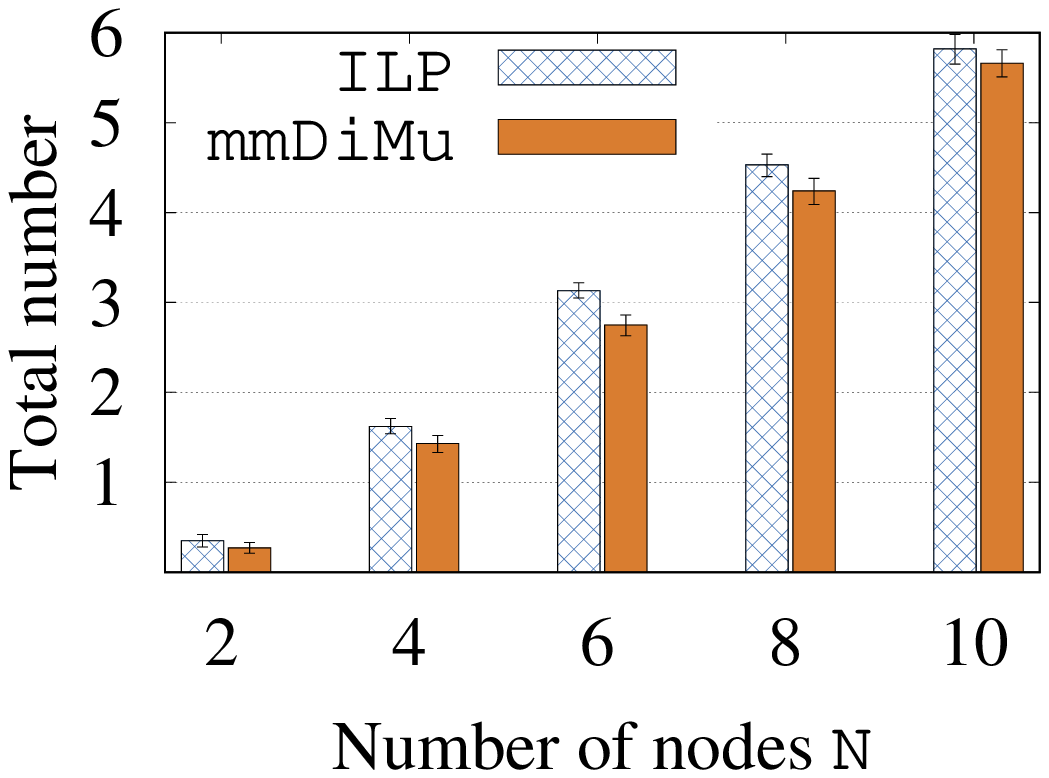}
		\vspace{-6mm}
		\caption{Relay transmission.}
		\label{fig:no_relay}
	\end{subfigure}
	\hspace{-1mm}
	\begin{subfigure}{0.23\textwidth}
    	\centering
    	\includegraphics[width=\columnwidth]{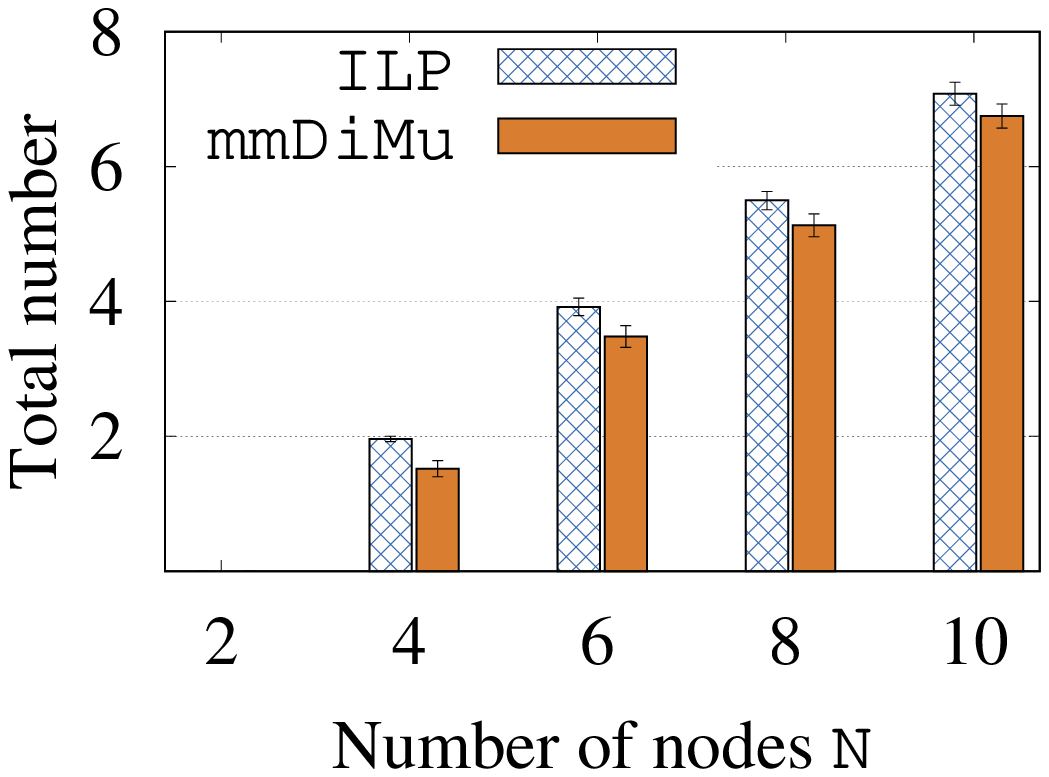}
		\vspace{-6mm}
		\caption{Concurrent transmission.}
		\label{fig:no_share}
	\end{subfigure} 
	\vspace{-2mm}
\caption{Total number of transmissions exploiting relay and spatial sharing for \texttt{ILP} and \texttt{mmDiMu} with $N = \{2, 4, 6, 8, 10\}$.}
\label{fig:stat_num}
\vspace{-5mm}
\end{figure}

As expected, \texttt{mmDiMu} performs worse than \texttt{ILP} because it autonomously schedules its transmission, disregarding the decision made by other PNs in the system. Let's consider the scenario in \fref{fig:sample_net} and the corresponding schedule in \tref{tab:example}. The completion time of \texttt{ILP} is $1$s lower than that of \texttt{mmDiMu}. Since \texttt{mmDiMu} sorts the nodes according to their SNR, the parent for $\footnotesize{\circled{4}}$ and $\footnotesize{\circled{5}}$ is $\footnotesize{\circled{3}}$, and $\footnotesize{\circled{4}}$ is served first. 
This results in $t^{\mathrm{s=3}} = 2$s. However, \texttt{ILP} is aware that scheduling node $\footnotesize{{\circled{5}}}$ first results in optimal completion time. As $N$ increases, the occurrence of this event increases as well. This reflects in the higher gain for \texttt{ILP} for larger $N$.

\noindent{\it Remark: The low complexity \texttt{mmDiMu} only requires $29.15\%$ additional completion time, in the worst case $N = 10$, as compared to \texttt{ILP}. 
Nevertheless, this additional time is significantly lower than that required by other algorithms.}

% ===========================================================================
%	Impact of relaying and spatial sharing
% ===========================================================================
\subsubsection{The importance of joint relaying and spatial sharing}
The substantial gain in the completion time demonstrated by our proposed algorithms (i.e., \texttt{ILP} and \texttt{mmDiMu}) emphasizes the importance of leveraging the relaying and spatial sharing gains jointly in mmWave multicast networks. To shed light on this aspect, \fref{fig:stat_num} and \fref{fig:stat_ratio} depict the number and ratio, respectively, of the relay and concurrent transmissions for \texttt{ILP} and \texttt{mmDiMu}. 
A transmission is a relay transmission if the transmitter is not the source node. 
A transmission pair is defined as a concurrent transmission if there is more than one transmission within the same time slot. For instance, in \tref{tab:example}, the number of relay transmission is $2$ (i.e., $\footnotesize{{\circled{3}} {\tiny\rightarrow} {\circled{4}}}$ and $\footnotesize{{\circled{3}} {\tiny\rightarrow} {\circled{5}}}$), and the number of concurrent transmissions is $2$ (i.e., $\footnotesize{{\circled{0}} {\tiny\rightarrow} {\circled{1}}~{\circled{2}}}$ and $\footnotesize{{\circled{3}} {\tiny\rightarrow} {\circled{5}}}$) for \texttt{ILP}.    

\begin{figure}[t!]
\centering
	\begin{subfigure}{0.23\textwidth}
    	\centering
    	\includegraphics[width=\columnwidth]{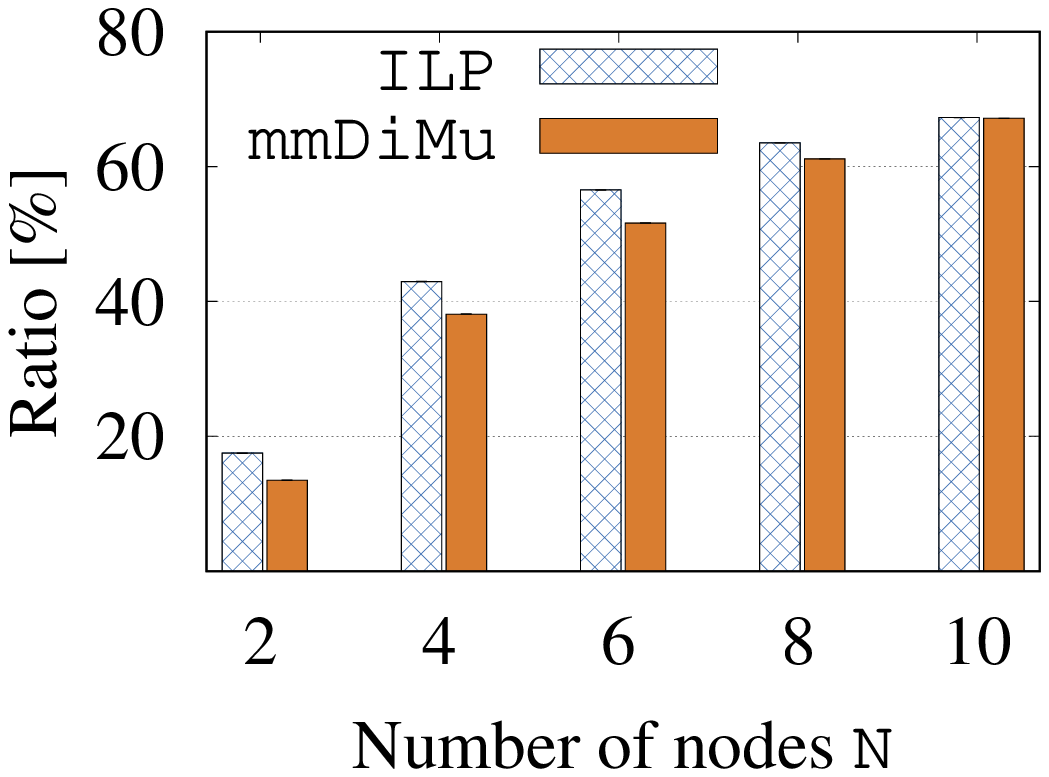}
		\vspace{-6mm}
		\caption{Relay transmission.}
		\label{fig:frac_relay}
	\end{subfigure}
	\hspace{-1mm}
	\begin{subfigure}{0.23\textwidth}
    	\centering
    	\includegraphics[width=\columnwidth]{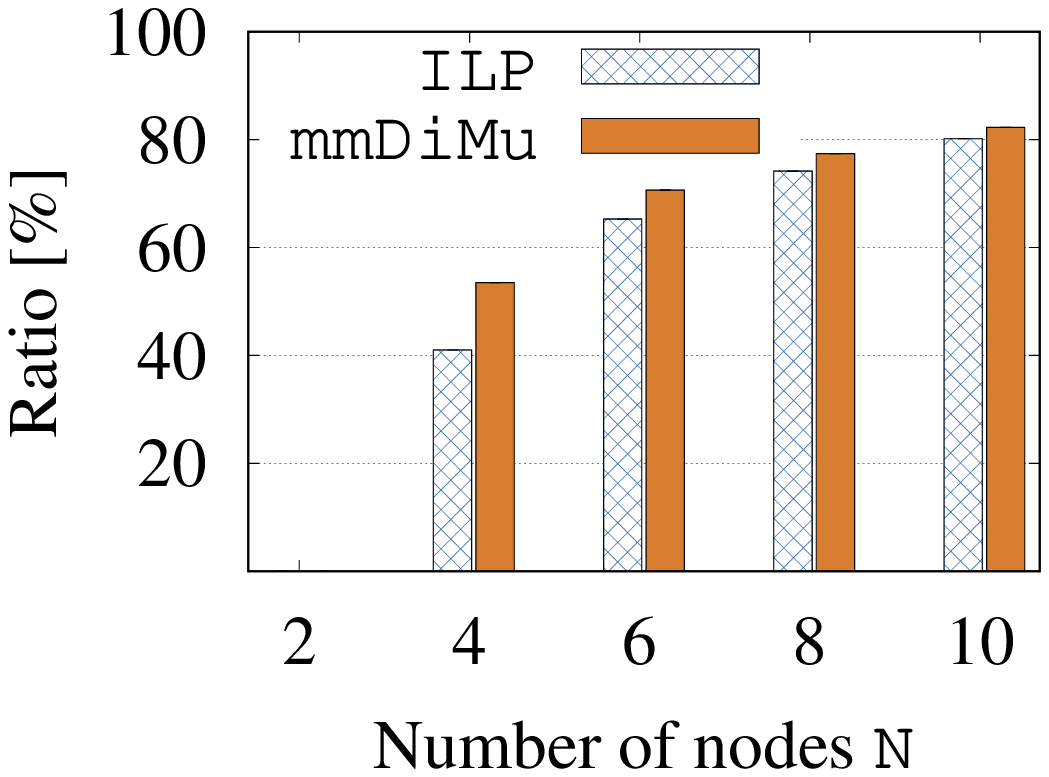}
		\vspace{-6mm}
		\caption{Concurrent transmission.}
		\label{fig:frac_share}
	\end{subfigure} 
	\vspace{-2mm}
\caption{The fraction of relay and concurrent transmissions for \texttt{ILP} and \texttt{mmDiMu} with $N = \{2, 4, 6, 8, 10\}$.}
\label{fig:stat_ratio}
\vspace{-2mm}
\end{figure}

\begin{table}[!t]
\scriptsize
\caption{An example of multicast scheduling for \texttt{ILP} and \texttt{mmDiMu} for the scenario in \fref{fig:sample_net}.}
\vspace{-2mm}
\centering
\begin{tabular}{|l|c|c|c|c|}
\hline
\multirow{ 2}{*}{Algorithm} & \multicolumn{2}{c|}{{\bf \texttt{ILP}}}		& \multicolumn{2}{c|}{{\bf \texttt{mmDiMu}}} 	\\ \cline{2-5}
							& Transmission link				& Time					& Transmission link 		& Time					\\ \hline\hline
\rowcolor{LightCyan}
time slot, $s=1$				& ${\tiny \circled{0}} \rightarrow {\tiny \circled{3}}$		& $\mathbf{1}$ 	& ${\tiny \circled{0}} \rightarrow {\tiny \circled{3}}$ & $\mathbf{1}$		\\ \hline 
\multirow{ 2}{*}{time slot, $s=2$}	& ${\tiny \circled{0}} \rightarrow {\tiny \circled{1}  \circled{2}}$	& $\mathbf{3}$	& ${\tiny \circled{0}} \rightarrow {\tiny \circled{1}  \circled{2}}$ &	$\mathbf{3}$		\\ \cline{2-5}
								& ${\tiny \circled{3}} \rightarrow {\tiny \circled{5}}$	& $2$	& ${\tiny \circled{3}} \rightarrow {\tiny \circled{4}}$ &	$1$		\\ \hline
\rowcolor{LightCyan}
time slot, $s=3$					& ${\tiny \circled{3}} \rightarrow {\tiny \circled{4}}$	& $\mathbf{1}$ 	& ${\tiny \circled{3}} \rightarrow {\tiny \circled{5}}$ &	$\mathbf{2}$		\\ \hline\hline
Completion time, $T$		& \multicolumn{2}{c|}{$5$sec}			& \multicolumn{2}{c|}{$6$sec}				\\ \hline
\end{tabular}
\label{tab:example}
\vspace{-5mm}
\end{table} 

In \fref{fig:stat_num}, the total number of relay and concurrent transmissions increases consistently with $N$. This increase is due to a higher communication diversity. 
We observe the total number of relay (in \fref{fig:no_relay}) and concurrent (in \fref{fig:no_share}) transmissions of \texttt{ILP} is consistently higher than that of \texttt{mmDiMu}. 
This indeed contributes to \texttt{ILP} outperforming \texttt{mmDiMu}.    
Firstly, \texttt{ILP} has a precise view of the entire network and knows the optimal policy; it first transmits to the nodes that can transmit with a high rate to another node later while maximizing spatial sharing gain. Unlike \texttt{ILP}, at each slot, each PN in \texttt{mmDiMu} opportunistically transmits to the CN set that maximizes the instantaneous sum throughput; the set of selected CNs is usually those that are located nearer to the PN. As a result, the CN set may not necessarily be the optimal set to relay the data to the remaining nodes at a later time. 
Secondly, each CN in \texttt{mmDiMu} only selects one PN. That said, a CN does not choose a secondary PN even if it potentially allows concurrent transmissions. 
As a result, this reduces the number of relay and concurrent transmissions of \texttt{mmDiMu}, and thus resulting in a higher completion time (as shown in  \fref{fig:var_n}). 

Further, we observe a high ratio of relay (up to $70\%$) and concurrent (up to $80\%$) transmissions over the corresponding total number of transmission for both \texttt{ILP} and \texttt{mmDiMu}. Precisely, a high number of concurrent transmission (in \fref{fig:no_share}) does not directly translate into a high number of the ratio (in \fref{fig:no_share}), but it highly depends on the total number of transmissions. This ratio confirms a large fraction of the performance gain roots from the exploitation of relaying and spatial sharing. 

\noindent{\it Remark: The gain achieved by \texttt{ILP} and \texttt{mmDiMu} mainly comes from the extensive exploitation of relaying and spatial sharing. This confirms the importance of leveraging these gains for mmWave multicast networks.} 

\begin{figure}[!tb]
	\centering
    \includegraphics[width=0.8\columnwidth]{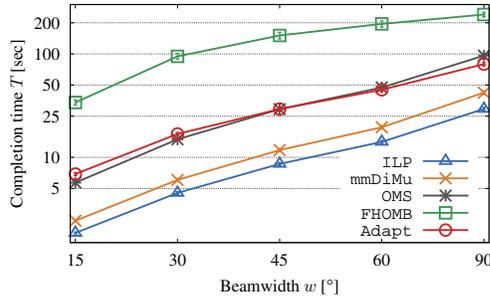}
    \vspace{-2mm}
	\caption{Completion time $T$ for different $w$. $N = 8$.}
	\label{fig:var_w}
	\vspace{-5mm}
\end{figure}

\begin{figure}[tb!]
\centering
	\begin{subfigure}{0.23\textwidth}
    	\centering
    	\includegraphics[width=\columnwidth]{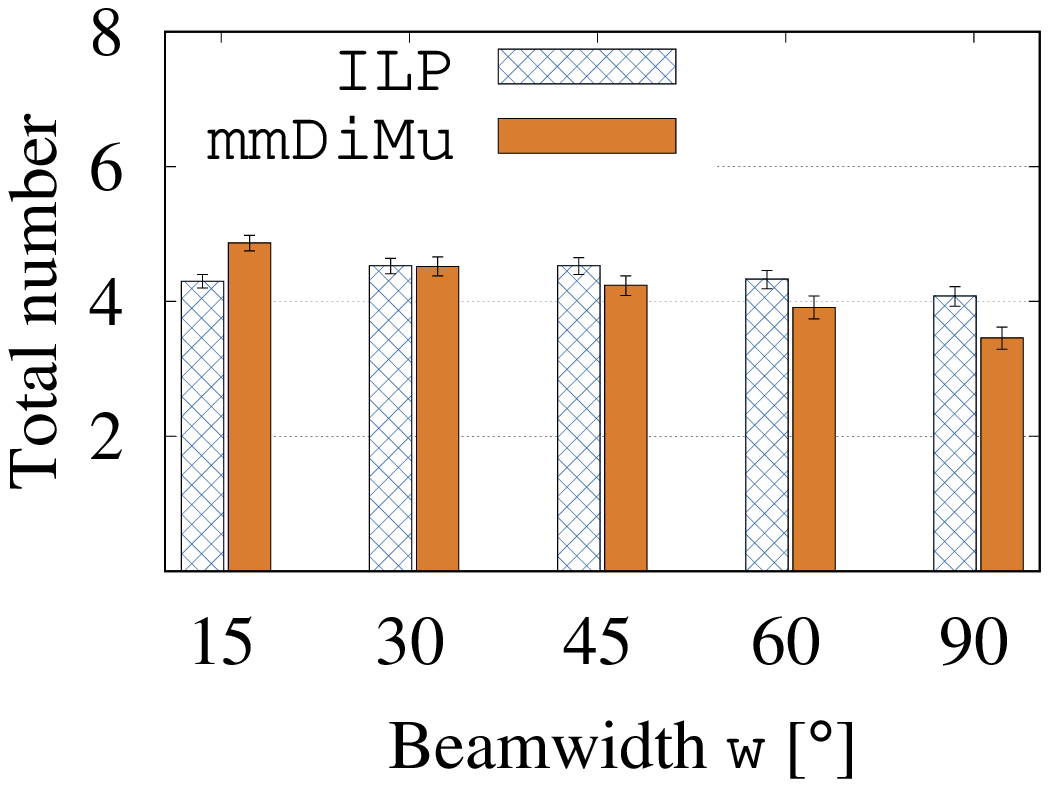}
		\vspace{-5mm}
		\caption{Relay transmission.}
		\label{fig:relay_w}
	\end{subfigure}
	\hspace{-1mm}
	\begin{subfigure}{0.23\textwidth}
    	\centering
    	\includegraphics[width=\columnwidth]{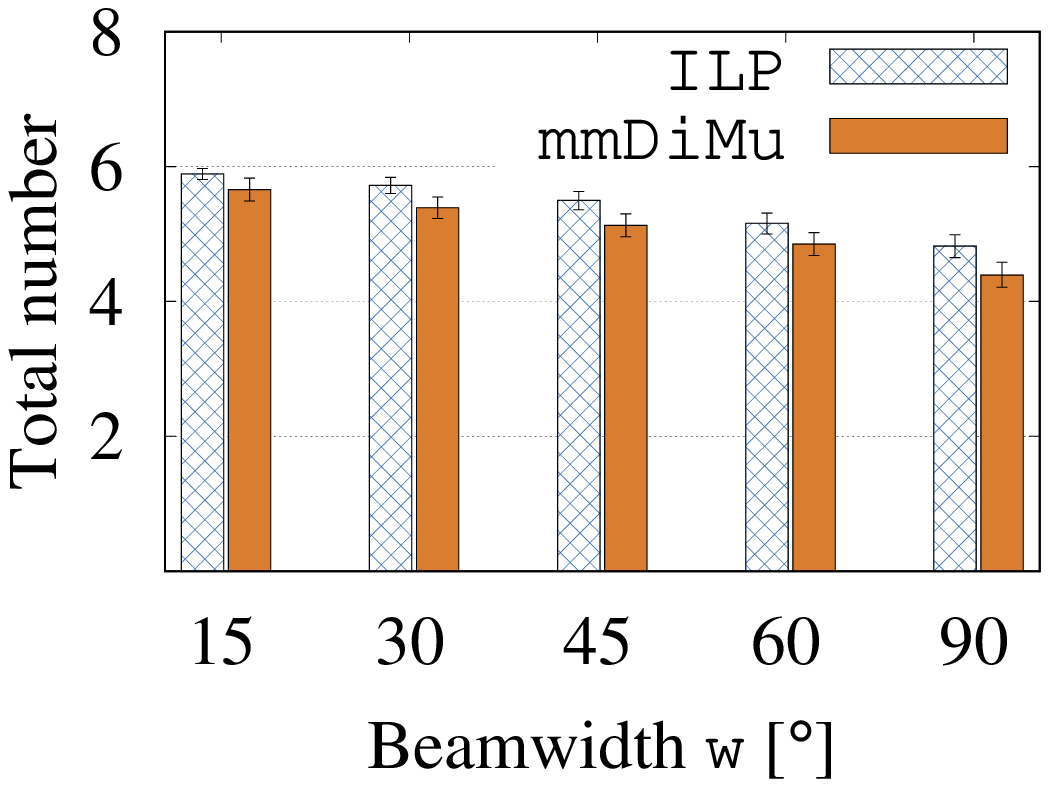}
		\vspace{-5mm}
		\caption{Concurrent tranmsission.}
		\label{fig:share_w}
	\end{subfigure} 
	\vspace{-2mm}
\caption{Total number of relay and concurrent transmissions for \texttt{ILP} and \texttt{mmDiMu} with $w = \{15^{\circ}, 30^{\circ}, 45^{\circ}, 60^{\circ}, 90^{\circ}\}$.}
\label{fig:share_relay_w}
\vspace{-5mm}
\end{figure}

\begin{figure}[!t]
	\centering
    \includegraphics[width=0.8\columnwidth]{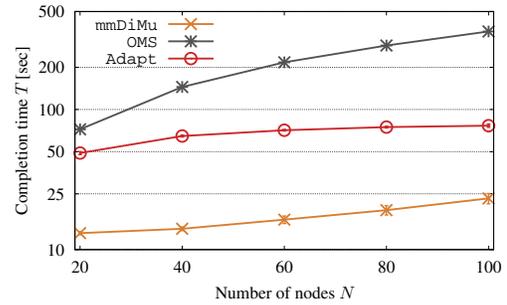}
    \vspace{-2mm}
	\caption{Completion time $T$ for $w = 45^{\circ}$. }
	\label{fig:n_var_w}
	\vspace{-5mm}
\end{figure}

\begin{figure}[!t]
\centering
	\begin{subfigure}{0.23\textwidth}
    	\centering
    	\includegraphics[width=\columnwidth]{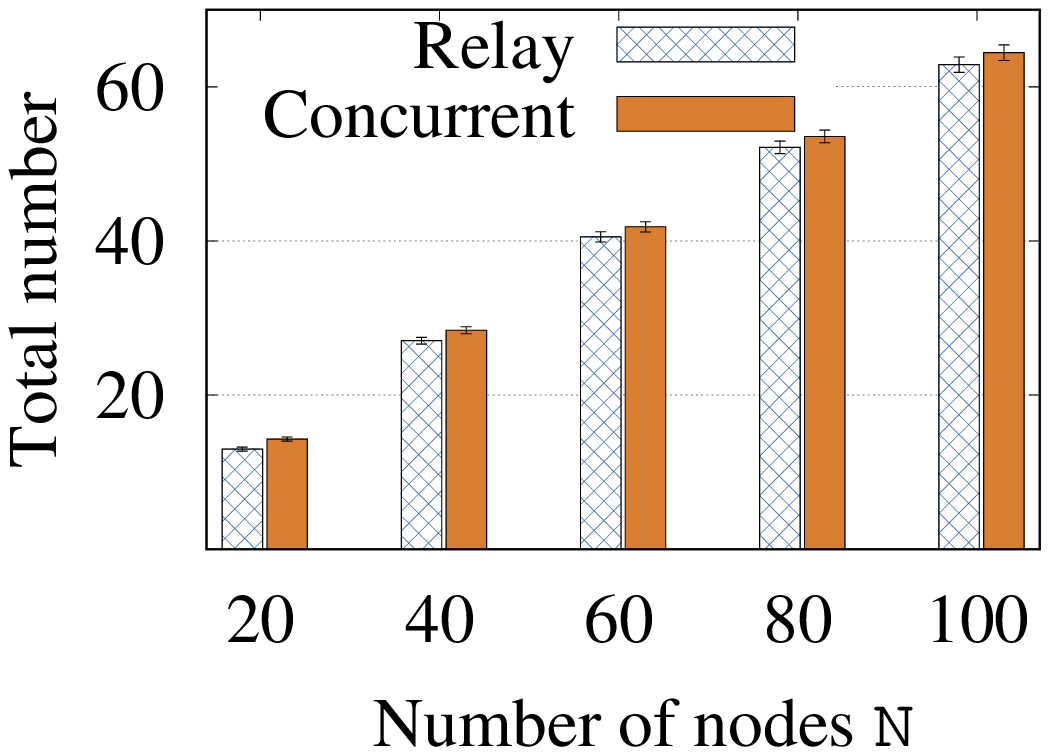}
		\vspace{-5mm}
		\caption{Number of transmissions.}
		\label{fig:relay_n}
	\end{subfigure}
	\hspace{-1mm}
	\begin{subfigure}{0.23\textwidth}
    	\centering
    	\includegraphics[width=\columnwidth]{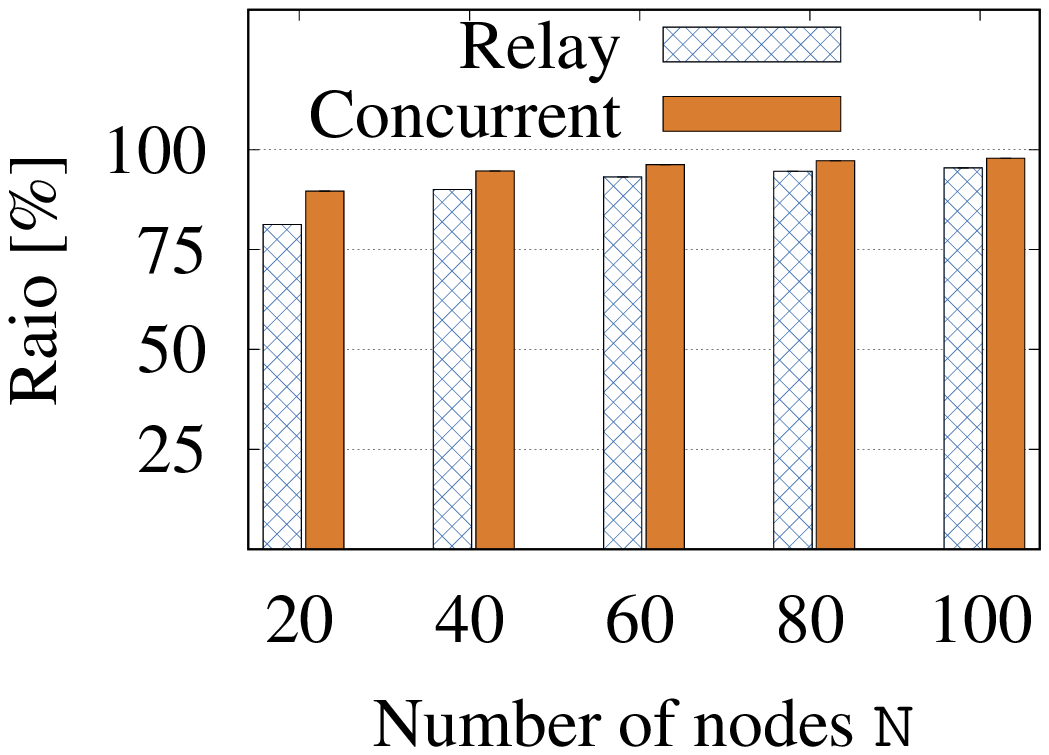}
		\vspace{-5mm}
		\caption{Ratio of transmission.}
		\label{fig:share_n}
	\end{subfigure} 
	\vspace{-2mm}
\caption{Total number and ratio for relay and concurrent transmissions for \texttt{mmDiMu} with $N = \{20, 40, 60, 80, 100\}$. }
\label{fig:share_relay_n}
\vspace{-5mm}
\end{figure}

% ===========================================================================
%	Impact of beamwidth
% ===========================================================================
\subsubsection{Impact of beamwidth $w$}
This section evaluates the impact of $w = \{15^{\circ}, 30^{\circ}, 45^{\circ}, 60^{\circ}, 90^{\circ} \}$ on the completion time $T$, while fixing the number of nodes $N = 8$.   

\fref{fig:var_w} shows a general trend that increasing beamwidth increases the completion time $T$. A wider beamwidth $w$ results in lower transmitting and receiving gains, which in turns results in a low data rate and a longer transmission time, and thus a high completion time $T$. 
This especially makes an impact on the algorithms that do not leverage relay. In particular, CNs located far away from the source node have to be served with very low transmission rates. 
Therefore, we observed an abrupt increase in the completion time of \texttt{OMS}, \texttt{FHOMB} and \texttt{Adapt}; the transmission time increases by $90.55$s, $206.20$s, and $72.74$s, respectively, as $w$ increases from $15^{\circ}$ to $90^{\circ}$. By manipulating relay, these CNs are reachable through a closer relay PN, resulting in a higher transmission rate. Therefore, the increase in transmission time for \texttt{ILP} and \texttt{mmDiMu} is lesser, i.e., only $15.26$s and $22.66$s, respectively, as $w$ increases from $15^{\circ}$ to $90^{\circ}$. 
Although the increase seems insignificant, it is still non-negligible. 
A wider $w$ improves the coverage area and a PN could cover more CNs. 
As a result, the number of relay and concurrent transmissions reduces, and the completion time increases. 
This is evident from the decreasing number of these transmissions as $w$ increases, as depicted in~\fref{fig:share_relay_w}. 

\noindent{\it Remark: Although a wider beamwidth increases the completion time, \texttt{ILP} 
and \texttt{mmDiMu} are less impacted by it, as compared to the other algorithms. }

\begin{figure*}[tb!]
\centering
	\begin{subfigure}{0.47\textwidth}
    	\centering
    	\includegraphics[width=\columnwidth]{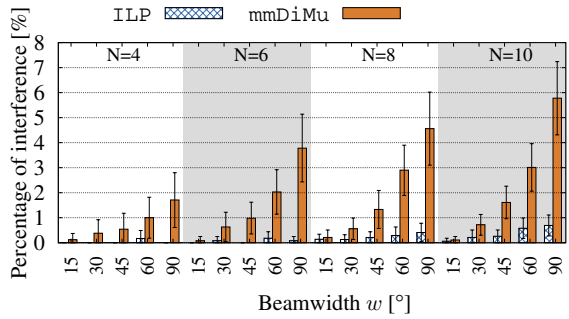}
		\vspace{-5mm}
		\caption{Omni-directional antenna receivers.}
		\label{fig:interference_omni}
	\end{subfigure} 
	\hspace{-4mm}
	\begin{subfigure}{0.47\textwidth}
    	\centering
    	\includegraphics[width=\columnwidth]{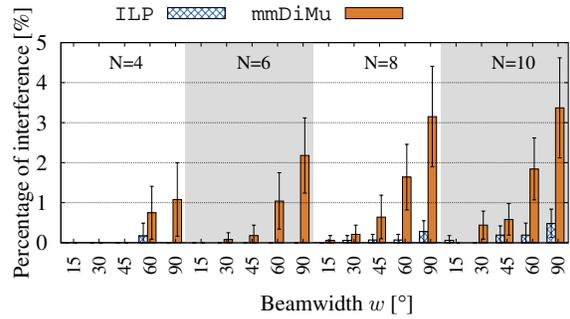}
		\vspace{-5mm}
		\caption{Directional antenna receivers.}
		\label{fig:interference_dir}
	\end{subfigure}
	\vspace{-2mm}
	\caption{Percentage of interference between concurrent transmission of \texttt{ILP} and \texttt{mmDiMu} for up to $N = 10$.}
\vspace{-5mm}
\end{figure*}

% ===========================================================================
%	Scalability
% ===========================================================================
\subsubsection{Scalability}
All previous results in this section only consider a maximum $N$ of $10$. This is due to the complexity and scalability issue of \texttt{ILP}. Nevertheless, it remains important as it provides insights on the algorithm performance difference to the optimal ones. 
Here, we demonstrate that even with a large $N$, our proposed \texttt{mmDiMu} algorithm achieves a significant gain as oppose to \texttt{OMS} and \texttt{Adapt}. \texttt{FHOMB} is removed from the comparison as it performs poorly even for cases with smaller $N$. 
We set $N = \{20, 40, 60, 80, 100\}$ with $w = 45^{\circ}$ for the following results. 

\fref{fig:n_var_w} depicts a significant increase in the completion time $T$ for \texttt{OMS}, but not for \texttt{Adapt}, as $N$ increases. 
As $N$ is large and continuously increases, the value of the lowest SNR does not change much, so does the transmit rate at each beam of the \texttt{Adapt} algorithm.
However, \texttt{OMS} greatly suffers from its opportunistic decisions. 
\texttt{mmDiMu} experiences less increment in completion time than \texttt{OMS}. Increasing $N$ increases the number of relay nodes and the opportunity of spatial sharing. This is evident from \fref{fig:share_relay_n}, where the number and ratio of relay and concurrent transmissions increase with $N$. 

\noindent{\it Remark:  
\texttt{mmDiMu} scales very well with the network density and achieves a significant reduction in the completion time as compared to \texttt{OMS} and \texttt{Adapt}. }

% ===========================================================================
%	Impact of interference
% ===========================================================================
\subsubsection{Impact of interference}
In theory, mmWave links mimic a pencil beam, and thus interference is negligible. However, the current off-the-shelf mmWave devices have a wider beam. In addition, the limitation in antenna design renders this assumption valid only in theory. 
Here, we evaluate the impact of transmit beamwidth $w=\{15^{\circ}, 30^{\circ}, 45^{\circ}, 60^{\circ}, 90^{\circ} \}$ for $N$ up to $10$ and $100$ for \texttt{ILP} and  \texttt{mmDiMu}, respectively, on the probability of mutual interference between concurrently transmitting pairs. 
We evaluate for interference characteristic for two type of antenna receiving modes: quasi-omnidirectional and directional,
% , as illustrated in \fref{fig:rx_mode}. 
In the first case, the receiver suffers from interference as long as it is within the beam's coverage of the transmitter. This type of receiving mode is as employed by default in the existing off-the-shelf devices (i.e., TP-Link Talon AD7200 multi-band wifi router \cite{adrouter}). 
In a very recent work on improving beam alignment in the mmWave device~\cite{Palacios:2018:ACO}, the authors are able to adaptively adjust the existing codebook available in the IEEE 802.11ad devices and optimize the beam pattern to obtain a higher directionality beam.
This shows the feasibility of implementing such receiving mode and thus it is important to also evaluate for interference when the receiver is in directional receiving mode.
In this case, to cause interference, not only that the receiver has to be within the beam coverage of the interfering transmitter, but the transmitter must also be within the coverage area of the receiving beam. 
\fref{fig:interference_omni} and \fref{fig:interference_dir} depict the percentage of mutual interference between the concurrently transmitting links as beamwidth $w$ increases for the respective type of receiver. 
Note that, due to the complexity of \texttt{OMS}, the interference percentage is only shown for up to $N=10$. 

\emph{Quasi-omnidirectional receiver: } 
\fref{fig:interference_omni} shows a general trend in which the percentage of interference increases with beamwidth $w$ and number of nodes $N$.  
As $w$ increases, so as the coverage area a transmitter, and thus increasing the probability of interfering the nearby nodes.  
As $N$ increases, so does the density of the network.
That said, the probability that one or more receiving node falling within the coverage area of a transmitter is higher, and thus the percentage of interference. 
Although increasing beamwidth causes higher interference, it only leads to a maximum percentage of interference of up to $6\%$ (see \fref{fig:interference_omni}) in the largest $N$ scenario; there is no interference in most scenario. 
Since the source of interference is due to the frequency of concurrent transmissions, \texttt{mmDiMu} experiences a slightly higher interference than \texttt{ILP}. 
\texttt{mmDiMu} indeed has a higher ratio of concurrent transmission as compared to \texttt{ILP} (refer \fref{fig:frac_share}). 
As shown in the example scenario in \fref{fig:tx_map}, \texttt{ILP} and \texttt{mmDiMu} has $7$ and $10$ total transmissions, respectively. 
Out of those, \texttt{ILP} and \texttt{mmDiMu} has $5$ and $9$ links, respectively, involved in concurrent transmission, which results a ratio of $71.43\%$ and  $90\%$, respectively. 
While \texttt{ILP} has no interference among the communication links, transmission from node {\scriptsize \circled{0}} of \texttt{mmDiMu} (see \fref{fig:tx_map_mmdemu}) causes interference to node {\scriptsize \circled{7}} when node {\scriptsize \circled{4}} transmits to {\scriptsize \circled{7}} simultaneously with $\scriptsize{{\circled{0}} {\tiny\rightarrow} {\circled{1}}}$. Nevertheless, even with omni-directional receiving mode, the percentage of interference in both algorithms is kept below $6\%$.       

\begin{figure}[t!]
\centering
	\begin{subfigure}{0.23\textwidth}
    	\centering
    	\includegraphics[width=\columnwidth]{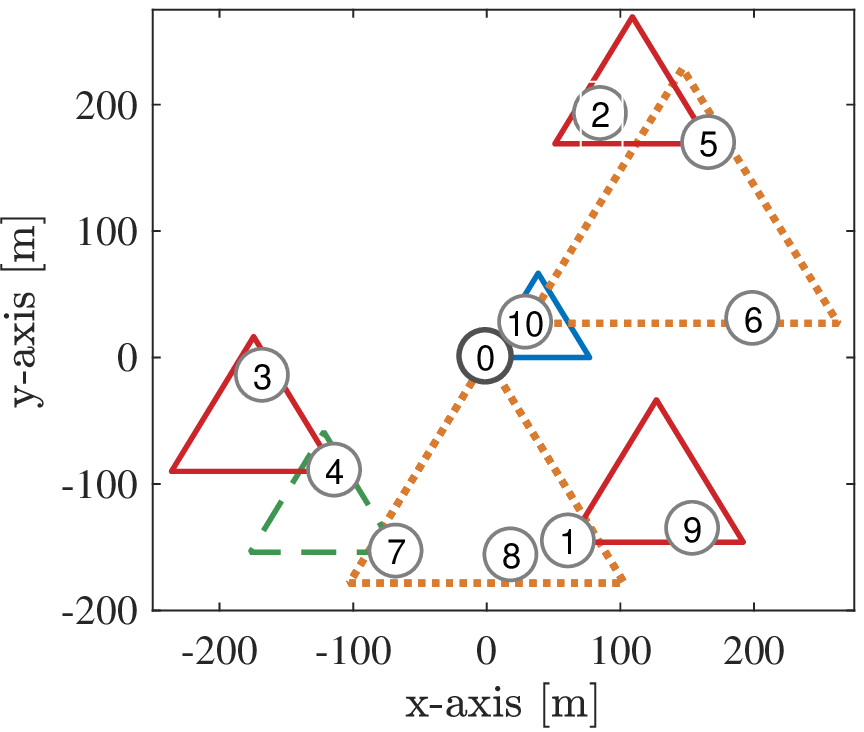}
		\vspace{-4mm}
		\caption{\texttt{ILP} with four slots.}
		\label{fig:tx_map_ilp}
	\end{subfigure} 	\begin{subfigure}{0.23\textwidth}
    	\centering
    	\includegraphics[width=\columnwidth]{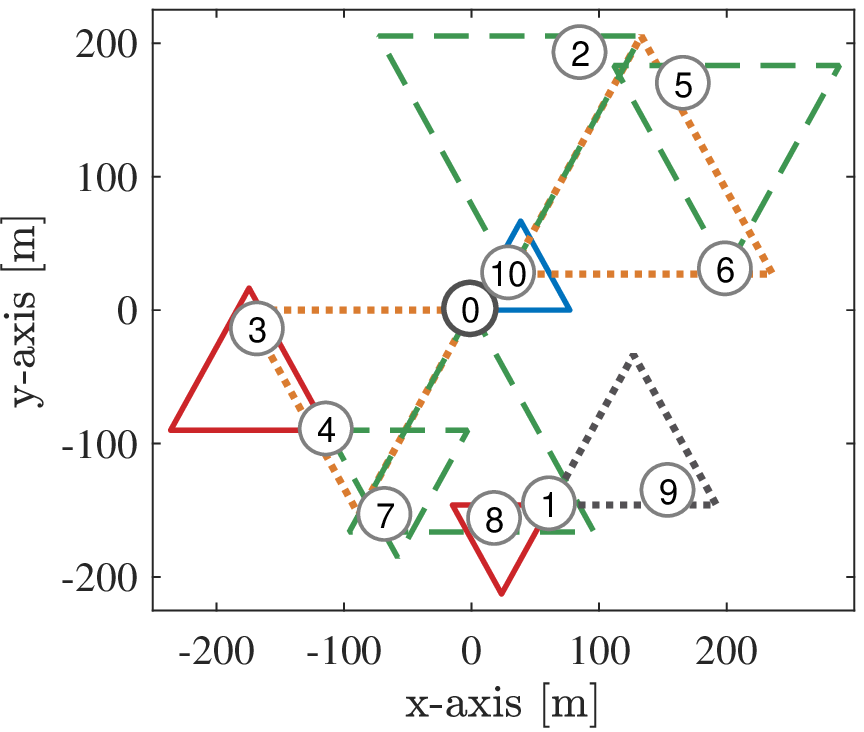}
		\vspace{-4mm}
		\caption{\texttt{mmDiMu} with five slots.}
		\label{fig:tx_map_mmdemu}
	\end{subfigure} 
	\vspace{-2mm}
\caption{Interference analysis for $N = 10$ and $w = 60^{\circ}$. The difference color represents the transmission at different slots.}
\label{fig:tx_map}
\vspace{-5mm}
\end{figure}

\emph{Directional receiver: }
When the receiver uses directional receiving mode, the percentage of interference becomes smaller (see \fref{fig:interference_dir}). 
This is due to the reason that interference only occur when the transmitter is within the beam of the receiver's beam, and the beam coverage is limited by the receiver's beamwidth $w$. For instance, while simultaneous transmission of $\footnotesize{{\circled{0}} {\tiny\rightarrow} {\circled{1}}}$ and $\footnotesize{{\circled{4}} {\tiny\rightarrow} {\circled{7}}}$ causes interference in the quasi-omnidirectional receiver's case, here, nodes {\footnotesize \circled{1}} and {\footnotesize \circled{7}} use directional reception, and thus avoiding interference from nodes {\footnotesize \circled{4}} and {\footnotesize \circled{0}}, respectively; the interfering nodes {\footnotesize \circled{4}} and {\footnotesize \circled{0}} are not within the directional receiving beam of nodes {\footnotesize \circled{1}} and {\footnotesize \circled{7}}, respectively. Therefore, we observed drops in the percentage of interference by up to $2.4\%$ (i.e., when $N = 10$, $w = 90^{\circ}$ for \texttt{mmDiMu}). 
The general performance trend is as seen in \fref{fig:interference_omni} for the same reasons explained above.  

\begin{figure}[!t]
    	\centering
    	\includegraphics[width=\columnwidth]{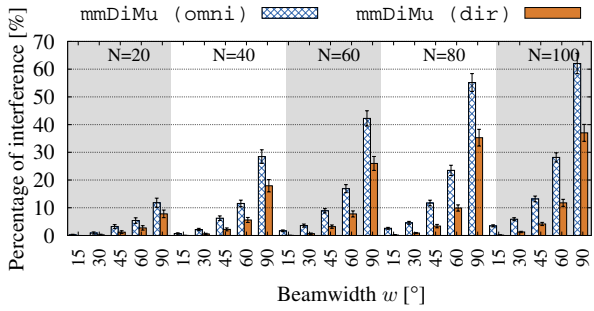}
	\vspace{-5mm}
\caption{Percentage of interference between concurrent transmission of \texttt{mmDiMu} for omni-directional (omni) and directional (dir) receiver.}
\label{fig:interference_dir_omni}
\vspace{-3mm}
\end{figure}

\emph{Scenario with $N$ up to 100: }
In \fref{fig:interference_dir_omni}, we show the interference's percentage of \texttt{mmDiMu} for up to $N = 100$ in order to provide some insight onto implementation setup for higher density scenarios. 
As seen, using directional receiving mode clearly provides a much lower percentage of interference for scenarios with higher density and beamwidth; the reduction is up to $25.01\%$ for $N = 100$ and $w = 90^{\circ}$. 
If the location of the receiver is known and the accuracy of beam alignment is high, using narrow beamwidth such as $15^{\circ}$ only has percentage of interference of up to $0.13\%$ in the worst case; many practical research work on mmWave use horn antenna with $w = 7^{\circ}$ \cite{sim:opprelay}.      

\emph{Remark: } 
Even for beamwidth as wide as $w = 45^{\circ}$, \texttt{mmDiMu} manages to keep the interference's percentage below $5\%$ for $N = 100$. 
We foresee future mmWave devices with highly directional and adjustable beam, in which, given any scheduling decision, the interference between the concurrently transmitting pairs in dense network can be further minimized.

\section{Discussion and Future Work}
\label{sec:discussion}

We dedicate this section to discuss the aspects that are out of scope of this paper yet are crucial to consider in developing a mmWave multicast scheduling algorithm. 

\subsection{Mobility}
The directional mmWave communication limits the coverage area. In this paper, we assume the nodes are static during the transmission period; a transmission period is on average equivalent to $1$s for transmitting a data frame of $1$Gbits. We deem this a valid assumption for a slow mobility network such as sports stadiums, concert halls, or urban vehicular scenarios with low speed cars. For networks with high-speed nodes such as high-speed train, uninterrupted connectivity can be ensured with features such as nodes tracking and beam switching. 

\subsection{Asynchronous Scheduling}
In this paper, we model the system as a slotted system with a synchronous slot, as shown in \fref{fig:sch_type} (which is based on the scenario in \fref{fig:sample_net}), where each transmission starts only at the beginning of each slot. 
Our proposed \texttt{ILP} algorithm is indeed designed to minimize the time difference between the simultaneously communicating pairs, but some small time gaps may persist. For instance, at time slot $2$ ($s=2$) in \fref{fig:sync}, $\footnotesize{{\circled{3}} {\tiny\rightarrow} {\circled{5}}}$ requires only $2$s to complete transmission, while $\footnotesize{{\circled{0}} {\tiny\rightarrow} {\circled{1}}~{\circled{2}}}$ requires $3$s. The time gap of $1$s at $\footnotesize{ \circled{3} }$ can potentially be used for another communication. As shown in \fref{fig:async}, $\footnotesize{{\circled{3}} {\tiny\rightarrow} {\circled{4}}}$ starts immediately as  $\footnotesize{{\circled{3}} {\tiny\rightarrow} {\circled{5}}}$ is completed. This way, the network transmission time is reduced by $1$s.  
While asynchronous scheduling improves the network transmission time, it results in a higher complexity algorithm. Therefore, the trade-off between gain and complexity must be considered carefully. 

\subsection{Scheduling Synchronization} 
In this paper, we assume that the scheduling decision is known by all the multicast nodes, and thus synchronization of transmissions among the nodes is feasible. However, the broadcast of scheduling information using mmWave is unreliable; mmWave is prone to blockages and suffers from high propagation loss. Therefore, the algorithm relies on information dissemination using the robust sub-6GHz transmissions. In fact, the protocol, namely fast session transfer (FST)\footnote{FST transfers the session between two physical channel to exchange information.}, that supports the coordination between mmWave and sub-6GHz interface for such purpose has already been outlined in the IEEE 802.11ad standard \cite{11ad:standard}. Specifically, the multicast nodes can exchange important scheduling information via sub-6GHz interface and the mmWave interface is dedicated for high rate data transmission only. In vehicular networks, such information can be exchanged via the robust dedicated short-range communication (DSRC) radio interface operating at 5.9GHz.  

\begin{figure}[!t]
\centering
	\begin{subfigure}{0.5\textwidth}
	\centering
    	\begin{tikzpicture}
			\draw (-4,-0.01) -- (4,-0.01);
			\draw [gray, thin] 	(-4,0.15) -- (-4,-0.15)
								(-2.4,0.15) -- (-2.4,-0.15)
								(-0.8,0.15) -- (-0.8,-0.15)
								(0.8,0.15) -- (0.8,-0.15)
								(2.4,0.15) -- (2.4,-0.15)
								(4,0.15) -- (4,-0.15);
			% 0 -> 3
			\filldraw [draw=none, fill=gcolor1] (-2.4,0) rectangle (-4,0.5);
			\node at (-3.65, 0.2) [circle, inner sep=1pt, minimum size=2.5mm, draw=gray!50, fill=white, font=\scriptsize]{0};
			\draw [->, thick] (-3.35, 0.2) -- (-3.05, 0.2);
			\node at (-2.75, 0.2) [circle, inner sep=1pt, minimum size=2.5mm, draw=gray!50, fill=white, font=\scriptsize]{3};
			% 0 -> 1,2
			\filldraw [draw=none, fill=gcolor2] (2.4,0) rectangle (-2.4,0.5);
			\node at (-0.7, 0.2) [circle, inner sep=1pt, minimum size=2.5mm, draw=gray!50, fill=white, font=\scriptsize]{0};
			\draw [->, thick] (-0.4, 0.2) -- (-0.1, 0.2);
			\node at (0.2, 0.2) [circle, inner sep=1pt, minimum size=2.5mm, draw=gray!50, fill=white, font=\scriptsize]{1};
			\node at (0.7, 0.2) [circle, inner sep=1pt, minimum size=2.5mm, draw=gray!50, fill=white, font=\scriptsize]{2};
			% 3 -> 5
			\filldraw [draw=none, fill=gcolor3] (0.8,0.525) rectangle (-2.4,1.125);
			\node at (-1.4, 0.825) [circle, inner sep=1pt, minimum size=2.5mm, draw=gray!50, fill=white, font=\scriptsize]{3};
			\draw [->, thick] (-1.1, 0.825) -- (-0.8, 0.825);
			\node at (-0.5, 0.825) [circle, inner sep=1pt, minimum size=2.5mm, draw=gray!50, fill=white, font=\scriptsize]{5};
			% 3 -> 4
			\filldraw [draw=none, fill=gcolor5] (4,0) rectangle (2.4,0.5);
			\node at (2.75, 0.2) [circle, inner sep=1pt, minimum size=2.5mm, draw=gray!50, fill=white, font=\scriptsize]{3};
			\draw [->, thick, white] (3.05, 0.2) -- (3.35, 0.2);
			\node at (3.65, 0.2) [circle, inner sep=1pt, minimum size=2.5mm, draw=gray!50, fill=white, font=\scriptsize]{4};
		\end{tikzpicture}
		\caption{Synchronous transmission}
		\label{fig:sync}
		\vspace{1mm} 
	\end{subfigure}
	
	\begin{subfigure}{0.5\textwidth}
	\centering
    	\begin{tikzpicture}
			\draw (-4,-0.01) -- (4,-0.01);
			\draw [gray, thin] 	(-4,0.15) -- (-4,-0.15)
								(-2.4,0.15) -- (-2.4,-0.15)
								(-0.8,0.15) -- (-0.8,-0.15)
								(0.8,0.15) -- (0.8,-0.15)
								(2.4,0.15) -- (2.4,-0.15)
								(4,0.15) -- (4,-0.15);
			% 0 -> 3
			\filldraw [draw=none, fill=gcolor1] (-2.4,0) rectangle (-4,0.5);
			\node at (-3.65, 0.2) [circle, inner sep=1pt, minimum size=2.5mm, draw=gray!50, fill=white, font=\scriptsize]{0};
			\draw [->, thick] (-3.35, 0.2) -- (-3.05, 0.2);
			\node at (-2.75, 0.2) [circle, inner sep=1pt, minimum size=2.5mm, draw=gray!50, fill=white, font=\scriptsize]{3};
			% 0 -> 1,2
			\filldraw [draw=none, fill=gcolor2] (2.4,0) rectangle (-2.4,0.5);
			\node at (-0.7, 0.2) [circle, inner sep=1pt, minimum size=2.5mm, draw=gray!50, fill=white, font=\scriptsize]{0};
			\draw [->, thick] (-0.4, 0.2) -- (-0.1, 0.2);
			\node at (0.2, 0.2) [circle, inner sep=1pt, minimum size=2.5mm, draw=gray!50, fill=white, font=\scriptsize]{1};
			\node at (0.7, 0.2) [circle, inner sep=1pt, minimum size=2.5mm, draw=gray!50, fill=white, font=\scriptsize]{2};
			% 3 -> 5
			\filldraw [draw=none, fill=gcolor3] (0.8,0.525) rectangle (-2.4,1.125);
			\node at (-1.4, 0.825) [circle, inner sep=1pt, minimum size=2.5mm, draw=gray!50, fill=white, font=\scriptsize]{3};
			\draw [->, thick] (-1.1, 0.825) -- (-0.8, 0.825);
			\node at (-0.5, 0.825) [circle, inner sep=1pt, minimum size=2.5mm, draw=gray!50, fill=white, font=\scriptsize]{5};
			% 3 -> 4
			\filldraw [draw=none, fill=gcolor5] (2.4,0.525) rectangle (0.8,1.125);
			\node at (1.15, 0.825) [circle, inner sep=1pt, minimum size=2.5mm, draw=gray!50, fill=white, font=\scriptsize]{3};
			\draw [->, thick, white] (1.45, 0.825) -- (1.75, 0.825);
			\node at (2.05, 0.825) [circle, inner sep=1pt, minimum size=2.5mm, draw=gray!50, fill=white, font=\scriptsize]{4};
		\end{tikzpicture}
		\caption{Asynchronous transmission}
		\label{fig:async}
	\end{subfigure}
\caption{The difference between synchronous and asynchronous scheduler.}
\label{fig:sch_type}
\vspace{-5mm}
\end{figure}
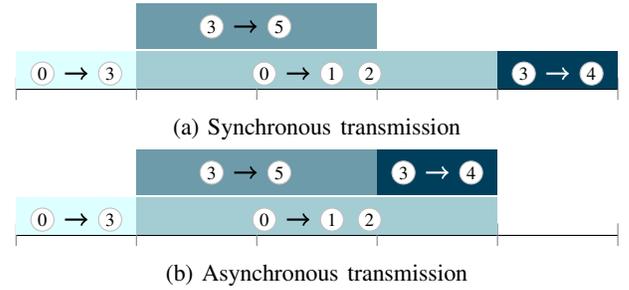

\subsection{Blockages}
A link is identified as a blocked link in the absence of either LOS or NLOS path. Based on the IEEE 802.11ad standard, this information could be obtained via nodes discovery phase upon the network initialization. In particular, during this phase, each communication pair performs beam training. When a transceiver pair fails to discover each other, the link between them is blocked.  While this is out of the scope of this paper, it can nevertheless be easily extended by removing a CN from its PN within the CNs matrix upon the identification of a blocked link. Proceeding with our algorithms, the blocked CNs will receive data only from a non-blocked relay PN.

\section{Conclusion}
\label{sec:conclusion}
 
In this paper, we investigate the challenge of multicasting in mmWave networks. We consider to jointly leverage \emph{relay transmission} to improve the reachability and link rate and \emph{spatial gain} by enabling simultaneous unicast and/or multicast communications. We formulate the problem with an ILP and provide a distributed solution called \texttt{mmDiMu}. The ILP solution generates optimal scheduling decisions while suffering from poor scalability. \texttt{mmDiMu} performs closely to the optimal and can scale to large networks with very dense settings due to its distributed nature. We show through extensive simulation that our proposed optimal \texttt{ILP} and distributed \texttt{mmDiMu} solutions provide significant gain over the multicast scheduling methods designed for sub-6GHz networks, in which we achieve up to $96.21\%$ reduction in completion time. Furthermore, in comparison with the adaptive beamwidth algorithm (namely \texttt{Adapt}) proposed for mmWave multicasting, we gain up to $78.22\%$ in completion time. Noteworthily, although interference reduction is excluded from the optimization objective, we achieve an impressively low (i.e., $5\%$) total  interference even with $45^{\circ}$ beamwidth in high-density network scenarios.  

There are still interesting open problems, such as studying the impact of user mobility, blockage, the tradeoff between efficiency, and complexity in asynchronous scheduling. We leave these for future work.

\bibliographystyle{IEEEtran}
\bibliography{ms}
\end{document}